\documentclass[12pt]{article}\usepackage[hyperfootnotes=false]{hyperref}
\usepackage{epsfig}
\usepackage{amsmath}
\usepackage{amssymb}
\usepackage{graphicx}
\newlength{\abstractwidth}
\renewcommand{\thefootnote}{\fnsymbol{footnote}}
\renewcommand{\thanks}[1]{\footnote{#1}} % Use this for footnotes
\newcommand{\starttext}{
\setcounter{footnote}{0}
\renewcommand{\thefootnote}{\arabic{footnote}}}
\renewcommand{\theequation}{\thesection.\arabic{equation}}
\newcommand{\be}{\begin{equation}}
\newcommand{\bea}{\begin{eqnarray}}
\newcommand{\eea}{\end{eqnarray}}
\newcommand{\beq}{\begin{equation}}
\newcommand{\ee}{\end{equation}}
\newcommand{\eeq}{\end{equation}}

\def\ba{\begin{eqnarray}}
\def\ea{\end{eqnarray}}

\def\12{{1 \over 2}}
\def\eq{&=&}

\def\ra{\rangle}

\def\simleq{\; \raise0.3ex\hbox{$<$\kern-0.75em
\raise-1.1ex\hbox{$\sim$}}\; }
\def\simgeq{\; \raise0.3ex\hbox{$>$\kern-0.75em
\raise-1.1ex\hbox{$\sim$}}\; }

\def\O2{\Omega_2}
\def\o{\omega}
\def\bi{\begin{itemize}}
\def\ei{\end{itemize}}

\def\sc{\setcounter{equation}{0}}

\def\W{$\Omega$}
\def\W'{$\Omega$}

\def\V{\Omega}
\def\V'{\Omega}

\def\o{{\cal{O}}}

\def\c{{\cal{C}}}

\def\bn{\bigskip \noindent}

\begin{document}
\renewcommand{\theequation}{\thesection.\arabic{equation}}

\begin{titlepage}
\rightline{}
\bigskip
\bigskip\bigskip\bigskip\bigskip
\bigskip
\centerline{\Large \bf {Computational Complexity and}}
\bigskip
\centerline{\Large \bf {  Black Hole Horizons}}

\bigskip

\begin{center}
\bf Leonard Susskind \rm

\bigskip

Stanford Institute for Theoretical Physics and Department of Physics, \\
Stanford University,
Stanford, CA 94305-4060, USA \\
\bigskip

\bigskip

\end{center}
\bigskip\bigskip
\bigskip\bigskip
\begin{abstract}

Computational complexity is essential to understanding the properties of black hole horizons. The problem of Alice creating a firewall behind the horizon of Bob's black hole is a problem of computational complexity. In general we find that while creating firewalls is possible, it is extremely difficult and probably impossible for black holes that form in sudden collapse, and then evaporate. On the other hand if the radiation is bottled up then after an exponentially long period of time firewalls may be common.

It is possible that gravity will provide tools to study problems of complexity; especially the range of complexity between  scrambling and exponential complexity.

\medskip
\noindent
\end{abstract}
\end{titlepage}

\starttext \baselineskip=17.63pt \setcounter{footnote}{0}

\tableofcontents

\sc
\section{ER=EPR and Computational Complexity}

A black hole may be highly entangled with a system that is effectively infinitely far away. Examples include an evaporating black hole after the Page time, when it has become entangled with its own cloud of Hawking radiation; and  the Thermofield-double (TFD) state \cite{Israel:1976ur}\cite{Maldacena:2001kr} of two non-interacting CFT's which can be interpreted as a pair of entangled black holes on disconnected spaces. Following \cite{Maldacena:2013xja}  In both cases I'll  assume that  the ER=EPR principle  \cite{Maldacena:2013xja} implies that the  black hole is connected to its purification by an Einstein-Rosen bridge (ERB).

Once we accept that entanglement creates Einstein-Rosen bridges, then it becomes possible for Alice's actions at one end to produce particles that come through the ERB,  and arrive at Bob's end. A sufficiently powerful blast of particles sent by Alice would constitute a firewall \cite{Almheiri:2012rt}\cite{Braunstein:2009my}. The mechanism was explicitly exhibited in \cite{VanRaamsdonk:2013sza}\cite{Shenker:2013pqa}.

The question that remains is this:

What are the conditions under which a disturbance at Alice's end will send a signal through the ERB? And how difficult  is it to accomplish? The ER=EPR duality is not enough  to tell us how difficult it is to send a signal through the ERB.  For that we need a second duality that also connects a geometric concept with an information-theoretic quantity.
This duality relates  \it distance from the horizon \rm  \ to \it computational complexity \rm \cite{Susskind:2013aaa}.
 Computational complexity is all about quantifying the degree of difficulty of carrying out a task. In this paper the task  is Alice's: to send a signal through the ERB to Bob\footnote{The signal cannot be received by Bob unless he passes through the horizon. Receiving a signal outside the black hole would violate locality and is forbidden by the non-traversability of wormholes.}.

Consider a quantum computer composed  of $K$ ``computational qubits."   The object of a computation is to start with a given state in the computational basis and carry out a particular unitary operation $U$ on the qubit system. Quantum computational complexity is a measure of how difficult it is to carry out the unitary operation. More precisely it is defined as the minimum number of  ``simple" unitary operations required to implement $U.$
By a simple unitary I mean a gate acting on some fixed number of qubits, let's say two.
A computation consists of applying a particular gate to a particular pair of qubits. From there the computation proceeds by picking another pair of qubits and applying another gate, and so on. The choice of qubits can be deterministic or statistical and the gates may be all the same, or updated at each stage. The complexity of $U$  is the minimum number of gates  required to implement $U.$

To understand the connection between complexity and black holes it is important to think of a black hole as an onion.
The near-horizon region of the black hole is a layered structure. Each layer has its own degrees of freedom. The CFT representation of the various layers differ in their degree of complexity.
 Operators supported on the outer edge of the near-horizon region---the onion skin---have the minimal complexity. As we peel the onion, the complexity increases. Eventually we get to the Planckian layer where the complexity has increased to a  particular value ${\c}_{\ast}$. We will derive the value of ${\c}_{\ast}$ later. But that's hardly the end of the story; the complexity can increase far beyond the Planckian value. This is closely related to the fact that Hawking radiation for old black holes  ``wells up" out of exponentially small distances.   I will argue that the right idea is not exponentially small-distances, but rather  exponentially large complexity.

\sc
\section{Strings and Qubits}

It is not necessary to have a string model for what follows. One can  assume that the relevant black hole  degrees of freedom are a system of qubits that interact in a way that makes the black hole a fast-scrambler \cite{Hayden:2007cs}\cite{Sekino:2008he}\cite{Lashkari}. The Hayden-Preskill circuit model  is an example. The point of the string model  is to connect the qubit model with dual gauge theory concepts \cite{Susskind:2013lpa}\cite{Susskind:2013aaa}.

For definiteness the formulas will be given for $(3+1)$-dimensional ADS at the Hawking-page (H-P) transition.
 If the dual gauge theory is regulated by a lattice regulator, each lattice cell has $N^2$ degrees of freedom where $N$ is the rank of the gauge group. Furthermore $N^2$ is also the entropy of a black hole at the H-P point. Thus a single lattice cell has enough degrees of freedom to describe an H-P black hole.

According to the UV/IR connection \cite{Susskind:1998dq} it should be possible to describe a unit H-P black hole by regulating the dual gauge theory to its bare minimum, i.e., a matrix theory describing  only the homogeneous degrees of freedom. One way to do that is to replace the system by a Hamiltonian lattice gauge theory \cite{Kogut} on a single lattice cell such as a spatial cube or a tetrahedron  If we choose the 't Hooft coupling to be of order unity, and let $N$ be large, the system has a thermal phase transition similar to the H-P transition. At the transition  a single  electric flux tube grows to length $N^2,$ as measured in lattice links or in ADS units. The resulting long string can be viewed as a model for a black hole along the lines of \cite{Susskind:1993ws}\cite{Sen:1994eb}\cite{Halyo:1996xe}\cite{Russo:1994ev}\cite{Peet:1995pe}\cite{Horowitz:1996nw}.

If one starts at an arbitrary location along the string, then at each site the string can continue along a small number of directions (two for the tetrahedron and also for the cube). Thus the quantum version of the string may be easily mapped to a system of $N^2$ qubits.  The state of the black hole can be modeled as a scrambled state of these qubits.

In this model the string scale $l_s$ and the ADS radius $l_{ads}$ are the same, and the thickness of the stretched horizon is also $l_{ads}.$ The Planck length is much smaller, related to the ADS scale by an inverse power of $N.$ (For $ADS(5) \times S(5)$ the power is $N^{-1/4}.$) Thus the stretched horizon is many Planck lengths thick.

The string can either be described as a system of computational qubits representing the way the electric flux changes direction at each site, or directly in gauge theory terms. Wilson-loops of various lengths are the most useful degrees of freedom to describe the gauge theory. A simple one-plaquette Wilson-loop  acts on the state of the string to locally change a few adjacent qubits. Such Wilson-loops are the closest thing to local gauge invariant operators in the dual lattice theory. They are also local on the long string.

In the continuum  gauge theory we can consider  an S-Wave local gauge invariant operator\footnote{Examples would include the energy-momentum tensor or the Lagrangian density. In the case of the energy momentum tensor the lowest angular momentum component would be $l=2.$ I am taking the liberty of calling it S-wave.} $\o.$  We usually think of these operators as living on the UV boundary of ADS, not the stretched horizon. But
 degrees of freedom which make up the entropy of the black hole have energy of order the Hawking temperature $T$ while the local boundary operators have infinite average  energy. To project out the relevant low-energy degrees of freedom we must integrate the local operator over a thermal time $\Delta t =T^{-1}.$ In the present case this means integrating over a time $l_{ads}.$  Such a time-averaged operator is described in the Schrodinger picture by referring it back to  $t=0,$

\be
W = \int_{\Delta t} dt \  \o(t) =  \int_{\Delta t} dt \ U(t) \o U^{\dag}(t)
\label{smear}
\ee

\bn
If we interpret $\o$ as the limit of a small Wilson-loop,  equations of motion allow us to express $W$ in terms of finite size Wilson-loops. Since the integral is over a limited time interval the Wilson-loops will not be very large. In fact they will be of ADS length. In other words they may be roughly  modeled by  single-plaquette operators in the coarse-grained lattice model.

Longer Wilson-loops that pass over the cell many times affect many qubits along the string. The result of evolving a single-plaquette Wilson-loop over a longer time interval  is to introduce longer loops into its Schrodinger picture representation. In a sense that I will make more precise, evolving increases the complexity of the Wilson loop.

In addition to ordinary Wilson loops we may consider decorated Wilson loops defined by replacing link variables by their time derivatives. This inserts electric operators conjugate to the link variables.

\subsection{Strings and Scrambling}

Consider the evolution of a Wilson loop $W$  that starts at time zero as a single-plaquette operator.

\be
W_p (t) = U(t) W U^{\dag}(t)
\label{precursor}
\ee

\bn
In the language of \cite{Susskind:2013lpa}\cite{Susskind:2013aaa} $W_p(t)$ is a precursor.
The subscript $p$ stands for precursor\footnote{I will use the term precursor for both the cases in which $t$ is positive and negative.}.

In the large $N$ limit the loop evolves into a linear superposition of loops in which the longest contribution grows exponentially with time. The reason  is that at any time  new contributions can ``bud" anywhere along its length. Thus the rate of growth is proportional to the length. After a time $t$ the longest contribution will have length,

\be
\exp{  \left(c  t/l_{ads}\right)}
\label{exp1}
\ee

\bn
where $c$ is  a numerical constant that depends on the 't Hooft coupling.

This behavior continues until the length of the longest contribution is the full length of the string, i.e., $N^2.$ This occurs at time,

\be
t_{\ast} \sim l_{ads} \log{N^2}
\ee

\bn
or in terms of the entropy $S,$

\be
t_{\ast} \sim l_{ads} \log{S}.
\label{tast}
\ee

\bn
Equation \ref{tast}  happens to define the scrambling time \cite{Sekino:2008he}.  Eq. \ref{tast} says that  it is the time  for the single-plaquette to evolve to a precursor involving the entire string. Note that this is true whether we evolve  to the future or to the past. Without being too precise at this point, we can say that the precursor gets more complex with time.

\sc
\section{Circuits and Complexity}

Quantum circuits were introduced into black hole physics by Hayden and Preskill \cite{Hayden:2007cs} as a simple substitute for intractable chaotic dynamics.

\subsection{Circuits}

The Hayden-Preskill  model is a quantum circuit that consists of $K$ qubits. In the case of the long string $K$ is given by the number of links,
which is also the rank of the gauge group and the entropy of the black hole,

\be
K = N^2 = S.
\ee

\bn
The number of qubits $K$ is called the width of the circuit.

The evolution is controlled by a universal set of two-qubit gates\footnote{Universal means that any unitary operator can be constructed by applying the universal gates. There are many examples of universal gates. The Kitaev-Solovay theorem says that they are in a sense interchangeable. }.  Hayden and Preskill consider two forms of evolution;  series, and  parallel. In the series circuit, at each step a single gate is allowed to act between some pair of qubits. The pair of qubits  may be chosen according to a rule or can be chosen at random. The following are assumed to hold \cite{Hayden:2007cs}:

\bn
1) Given any unitary operator $U$ it may be implemented by a series circuit with no more than an exponential (in $K$) number of gates.

\bn
2) Scrambling can be accomplished with a number of gates of order $K \log{K}.$ This does not have the status of a theorem, but fits very well with what is known from black hole scrambling \cite{Hayden:2007cs}\cite{Sekino:2008he}\cite{Lashkari}.

\bn

In simulating Hamiltonian dynamics of a system of degrees of freedom, the parallel circuit is the better model. At each discrete time  the qubits are paired randomly, and $K/2$ gates are allowed to simultaneously act. Every qubit interacts once in each time-interval, but with a randomly chosen partner. The number of such parallel time-steps is called the depth of the circuit.

In the parallel case statement 2) can be expressed as follows:

\bn

2')  Scrambling can be accomplished with a circuit of depth $D$ where

\be
D = \log{K}.
\label{depth}
\ee

\bn
Scrambling can be understood \cite{Hayden:2007cs} in circuit terms by an intuitive picture which closely resembles the string argument in the previous section. Consider a particular qubit. After one time interval that qubit will have interacted with just one other qubit. After two time intervals it will have \it indirectly \rm  \ interacted with four qubits, and after $n$ time intervals, with $2^n$ qubits. The scrambling time is roughly the time for any given qubit to have made indirect contact with every qubit; in other words

\be
2^n =K.
\label{exp2}
\ee

\bn
or

\be
n \sim \log{K}
\ee

\bn
The exponential growth in \ref{exp2} is very similar to the exponential growth of Wilson loops in \ref{exp1} and also of the way diffusion spreads over the horizon of a black hole.

Comparison between black hole scrambling and circuit scrambling suggests that a black hole may be thought of as a quantum circuit of $K=S$ qubits, executing one parallel operation every ADS time interval. Equivalently it executes individual gates at a rate given by the product of the entropy and temperature.

\be
\rm rate \ of \ computation \it \ = ST.
\label{rate}
\ee

\subsection{Complexity}

A quantum computation consists of applying a given unitary operator to a $K$-qubit system. The complexity of the task, called $\c,$ can be simply defined as \it the minimum number of gates that it takes to implement the unitary operator. \rm  \  In the case of the parallel circuit the number of gates is the product of the width and the depth. By definition it increases linearly with time; i.e., with the depth of the circuit. But that does not necessarily mean that the complexity of the output increases linearly. The complexity is defined as the minimum number of gates needed to do a job; not the number of gates in some particular setup.

Nevertheless for the Hayden-Preskill type circuits I will assume  that over some range of time the complexity does increase linearly. The range cannot be infinite, since as we will see, complexity is bounded, but the time for saturating the bound is exponentially large. On even larger time scales the complexity is quasi-periodic with an average period equal to the doubly exponential quantum recurrence time.

The depth of a circuit is the length of time that the circuit runs in some suitable time units. In \cite{Sekino:2008he} the time-step was argued to be the inverse energy-per-qubit or the inverse temperature. Thus the
complexity of  $U$ is the product of the number of qubits $K,$ and the minimum number of time steps of duration $l_{ads},$ that it takes to implement $U.$

 \be
 \c =  tTK.
 \ee

\bn
and for the case of the gauge theory at the H-P point,

\be
\c = N^2 t/l_{ads}
\ee

\bn
The complexity needed to scramble \cite{Hayden:2007cs}\cite{Sekino:2008he} is

\bea
\c_{\ast} \eq  K \log{K} \cr \cr
\eq N^2 \log{N^2} \cr \cr
\eq S \log{S}
\label{cstar}
\eea

\bn
corresponding to a running time,

\be
t_{\ast} = l_{ads} \log{S}.
\ee

\bn

From either the circuit model or the string model it is clear that up until the scrambling time the complexity  increases by a simple straightforward mechanism:  the increasing  size of  systems   that have been in indirect contact. Scrambling occurs when all qubits have been in indirect contact.

\bn

 However, \ref{cstar} is by no means the maximum complexity that can be achieved. It is very far from saturating the maximum for $K$ qubits. If the circuit continues to run, the complexity will  continue to increase until it reaches a maximum. The existence of a maximum follows from the fact that any unitary can be implemented by a circuit with no  more than an exponential  number of gates (exponential in $K$)\footnote{It can be shown that almost all unitary matrices require an exponential number of gates \cite{Knill}.}. Thus

$$
\c_{max} \sim e^{K}
$$

\bn

The time to achieve this degree of complexity is of order the  classical recurrence time. Once this much time has elapsed the complexity stops increasing, although it may fluctuate.

Finally, consider the complexity of precursor operators such as \ref{precursor}. The complexity of the one-plaquette operator $W(0)$ is very small; in the qubit description of the  string, a single-plaquette operator acts on a small number of qubits, so we can be sure that  a small number of gates will suffice to implement it. More generally a precursor \ref{precursor} can  be implemented by applying $U^{\dag},$ $W,$ and $U$ in succession. Therefore the complexity is bounded from above by a value
$\sim N^2 { t / l_{ads} }.$

I will assume that the lower bound is of the same order so that the complexity of a precursor is

\bea
\c &\approx& N^2 t/l_{ads} \cr \cr
\eq  \frac{t}{l_{ads}}K \cr \cr
\eq \frac{t}{l_{ads}}S.
\label{c = K t/lads}
\eea

\bn
One might wonder
\begin{figure}[h!]
\begin{center}
\includegraphics[scale=.3]{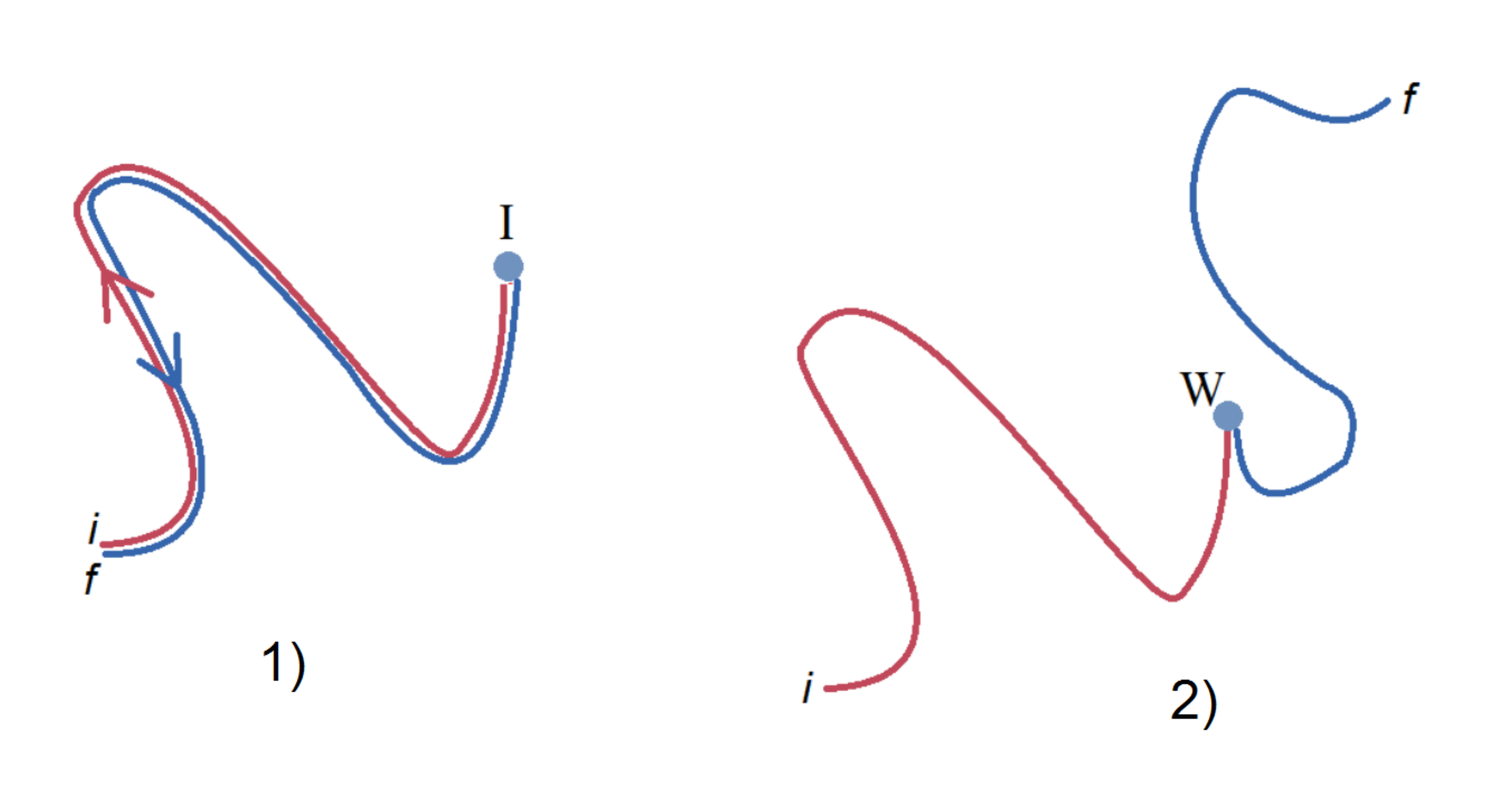}
\caption{1) A  system is run forward from  a state $i$ for a time $t.$ The sign of the Hamiltonian is flipped and the system evolves for another time $t.$ At the end the final state $f$ is the same as the initial state. The symbol $I$ indicates that nothing was done at the end of the first interval to disturb the configuration.  \ 2) The operator $W$ is inserted before the evolution is reversed. }
\label{if}
\end{center}
\end{figure}
whether the complexity might be much smaller due to a cancelation between $U$ and $U^{\dag}.$
For example, suppose that in \ref{precursor}  $W$ is replaced by the unit operator. Then no matter how large $t$ is, the result has zero complexity since $U(t)$ and $U^{\dag}(t)$ cancel to give the unit operator. There is a classical analog shown in figure \ref{if} - 1). A  (red) trajectory begins at point $i$ and evolves for a time $t$ to point $I.$ At that point, the Hamiltonian is reversed in sign, and the trajectory is continued (blue) for another time $t,$  to the final point $f.$ If no inaccuracy is introduced then the blue trajectory will exactly  re-trace the red, and $f$ will be the same point as $i.$ The quantum analog is that the  red plus blue trajectory is replaced by $U(t) U^{\dag}(t) = I.$

 Now let $I$ be replaced by $W.$ Should we expect similar cancelation? The answer is no; one can see the point classically.
The classical analog  of inserting $W$ is to introduce a   perturbation at the end of the red trajectory, and then continue on with the opposite sign hamiltonian. Approximate cancelation would mean that the final point is close to the initial.
 However this is not expected because of the chaotic nature of the system. The action of $W$ can be thought of as a small (classical) error\footnote{See appendix A}, but small errors don't stay small in a chaotic system. As shown in figure \ref{if} -2), after a very short time chaos will cause the blue trajectory to separate from the red and go to an entirely different final point $f.$ Therefore the cancelation which occurred when we replaced $W$ by the unit operator will not even approximately occur.

\bigskip

The vast range of complexity between scrambling and recurrence is very subtle. It has a different origin than complexity below the scrambling limit. Complexity smaller than \ref{cstar} is associated with the size of clusters of qubits that have been in contact. This saturates at size $K.$ Beyond that, the increase of complexity is less intuitive. Operators of larger complexity are sums of products of qubits, or sums of Wilson loops. The increasing complexity is not due to the size of such products growing; it has more to do with the the growing number of terms in the sums\footnote{A technical point explained by Hayden and Preskill \cite{Hayden:2007cs} is that the circuit produces a 2-design at the scrambling time. With increasing complexity it produces t-designs of increasing $t.$ At the classical recurrence time the output unitary will be Haar-random.}.

\sc
\section{ Geometry   and Complexity }

\subsection{The Layered Stretched Horizon}

We will now come to the duality between complexity and geometric location in the near-horizon region\footnote{This should not be confused with the geometrization of complexity in \cite{Nielsen}.}.

\begin{figure}[h!]
\begin{center}
\includegraphics[scale=.3]{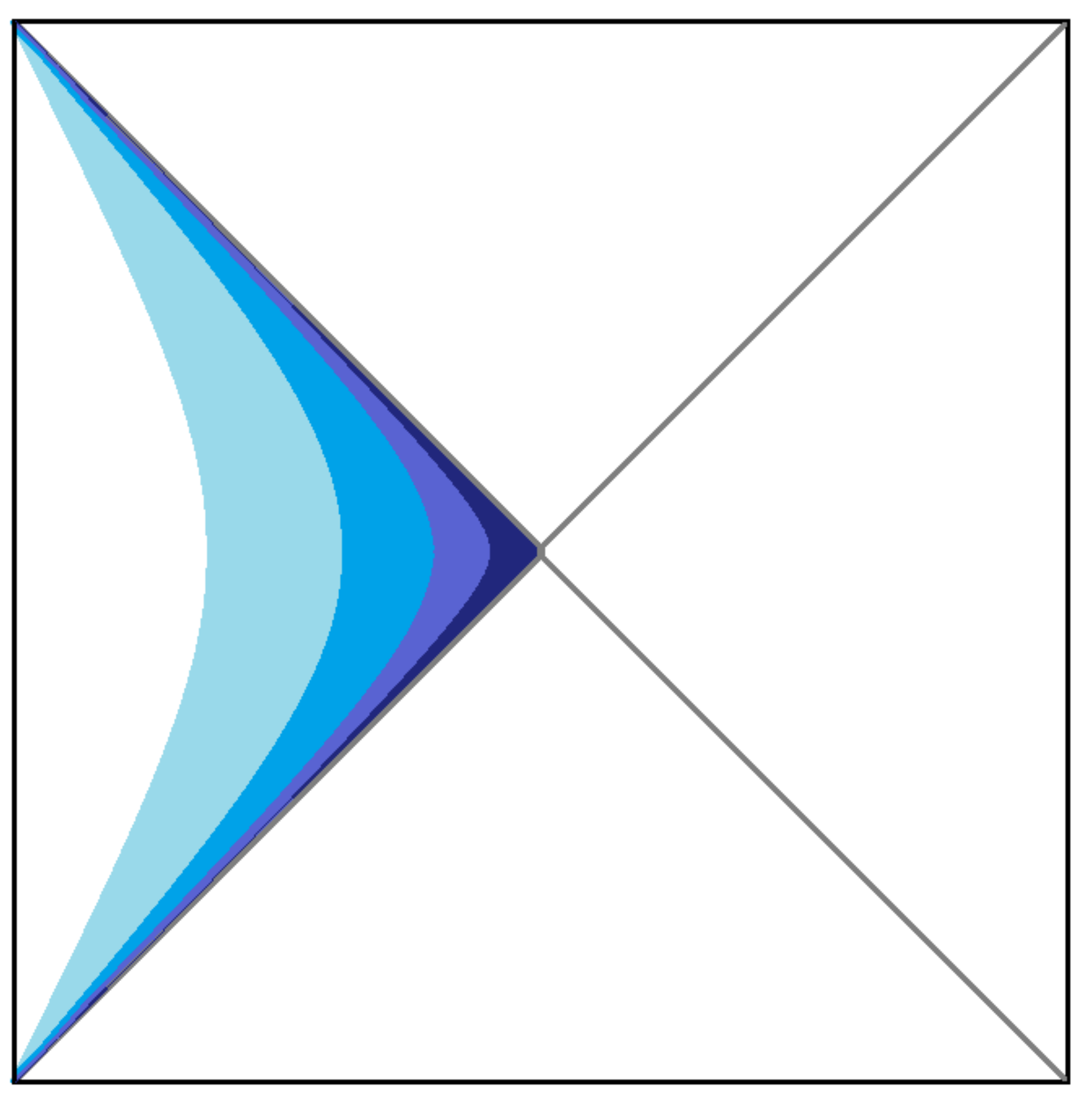}
\caption{Penrose diagram showing the layered structure of the near-horizon region of an ADS black hole. }
\label{e1}
\end{center}
\end{figure}

In the 't Hooft limit with $g^2N \sim 1$ the string and Planck scale are parametrically different, with the string length being much larger than the Planck length; in fact the string scale is of the same order as the ADS radius of curvature.
For this reason the stringy stretched horizon is much thicker than a Planck length, and we can consider it to be an onion-like structure with many layers as shown in figure \ref{e1}. The outer boundary of the stretched horizon is roughly an ADS length above the true horizon.

\begin{figure}[h!]
\begin{center}
\includegraphics[scale=.3]{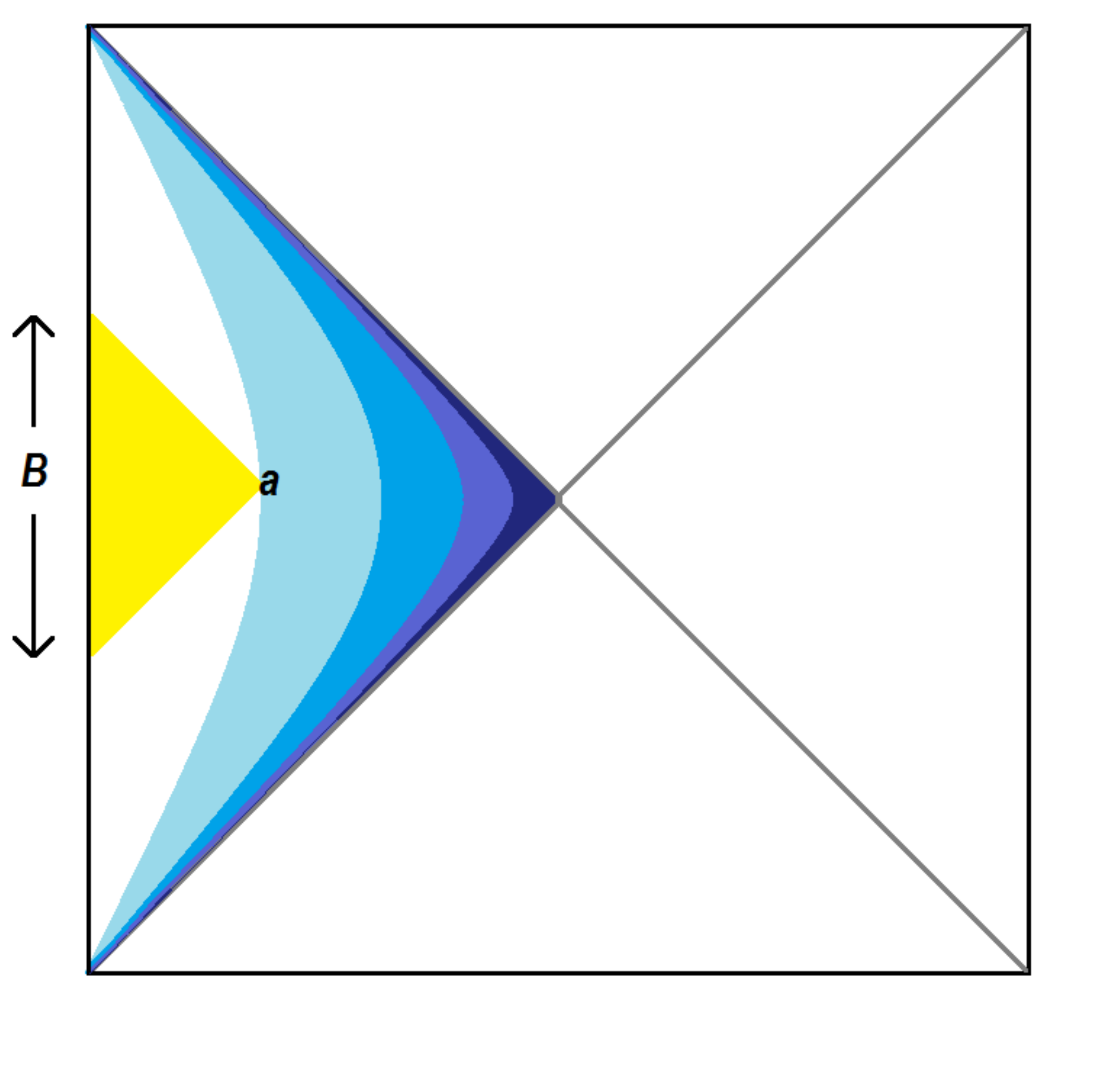}
\caption{A degree of freedom at the outer boundary of the stretched horizon may be represented in terms of boundary degrees of freedom along the vertical base of the yellow triangle. }
\label{e2}
\end{center}
\end{figure}

The various layers can be identified with varying degrees of complexity \cite{Susskind:2013aaa}.  Consider a  bulk operator $\bf{a},$   localized on the outermost layer, as in figure \ref{e2}.

The operator $\bf{a},$ is assumed to be unitary. For example it could be an complex exponential of  local field at $\bf{a}.$  To associate a degree of complexity with  $\bf{a}$ we may express it in terms of boundary operators using the   Hamilton, Kabat, Lifschytz, Lowe construction \cite{Hamilton:2005ju}. The vertical base $\bf{B}$ of the large yellow triangle in figure \ref{e2} shows the region of the boundary that is involved in the construction of $\bf{a}.$ In practice the most important contributions come from the ends of the base since they are light-like to $\bf{a}.$

To represent $\bf{a}$ as a boundary CFT operator in the Schrodinger picture, the local operators on $\bf{B}$ should be run back to $t=0,$ which replaces them by non-local Wilson-loops. Given that the time interval represented by $\bf{B}$ is one ADS length, the Wilson-loops in the regulated lattice gauge theory will be dominantly single-plaquette operators. Therefore the operator $\bf{a}$ is minimally complex, with a complexity of order

\be
\c_a \sim 1.
\ee

\bn
Now let us proceed inward using the same logic.
 Let $l_a$ be the distance of the point $\bf{a}$ from the horizon.
From figure \ref{e2} and a bit of elementary geometry
\begin{figure}[h!]
\begin{center}
\includegraphics[scale=.3]{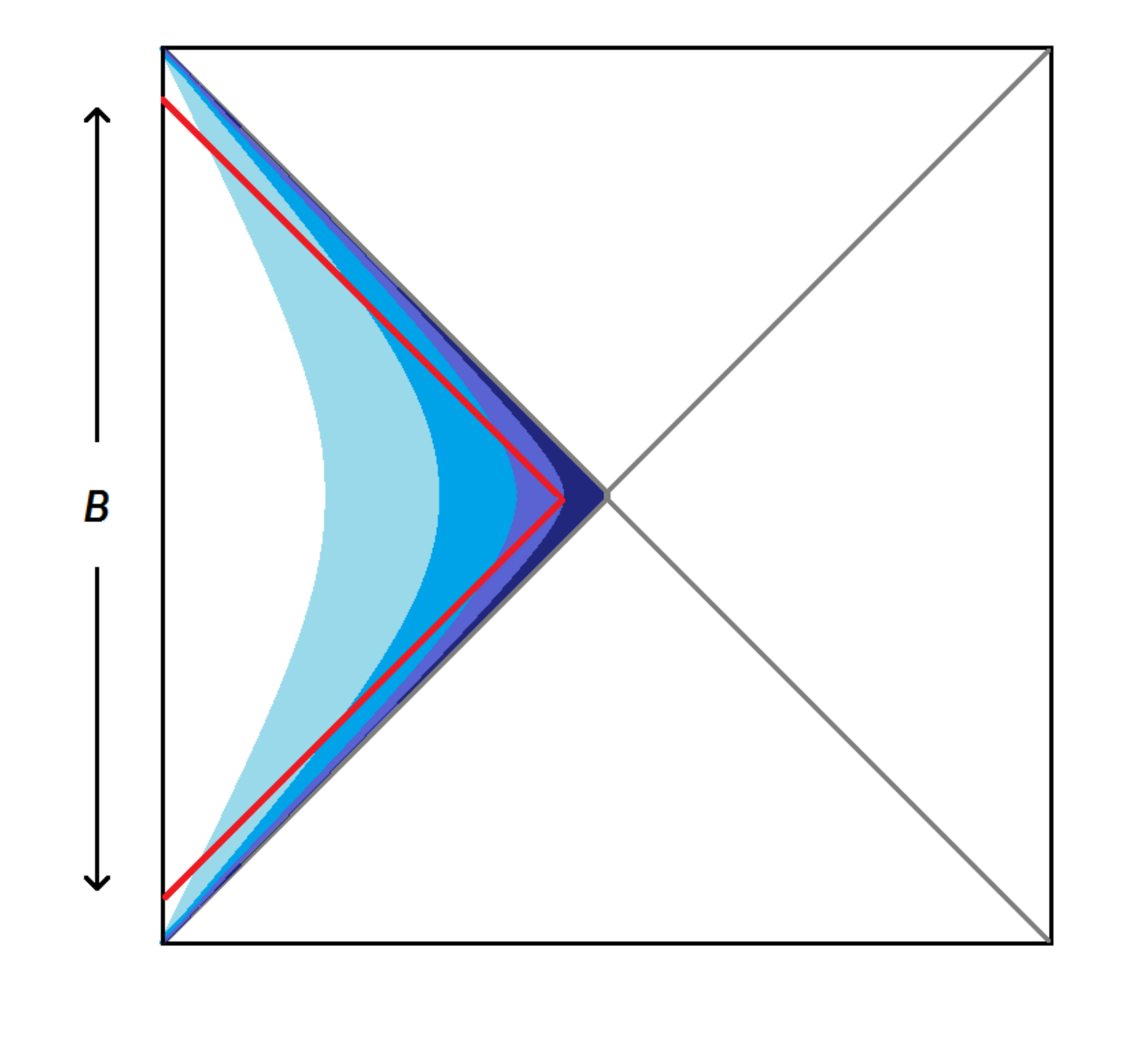}
\caption{A degree of freedom at the Planck distance from the horizon is represented in terms of boundary degrees of freedom
between the red lines. }
\label{e3}
\end{center}
\end{figure}
 one can see that the time-interval along the base of the yellow triangle  scales like

\be
t=l_{ads} \log{\frac{l_{ads}}{l_a}}
\label{logtime}
\ee

\bn
for small $l_a.$

As before we  assume the dominant contributions come from the endpoints of the base of the triangle.

Thus if we combine \ref{logtime}   with \ref{c = K t/lads} we find,

\be
\frac{l_a^2}{l_{ads^2}} = e^{-\c / S}
\label{fundformula}
\ee

\bn
This is a fundamental  relation between geometry (manifested as distance from the horizon), complexity, and entropy.  Note that the quantity in the exponent is the complexity per bit of entropy or complexity per qubit.

Let us apply \ref{fundformula} to a point $\bf{a}$ a Planck distance $l_p$  from the horizon. Recalling that the entropy of a H-P black hole is proportional  to $(l_{ads}/l_p)^2,$ the formula becomes,

\be
 S = e^{\c / S}
\label{S = exp c over s}
\ee

\bn
 or, solving for the complexity,

 \be
 \c = S \log{S}.
 \label{c=slogs}
 \ee

\bn
Notice    that this is also the scrambling complexity \ref{cstar}. This  implies a new information-theoretic meaning to the Planck length. The usual perspective is in terms of the Bekenstein-Hawking entropy formula: the entropy of a horizon is proportional to its area in Planck units. From this point of view area is primary. The new perspective focuses on complexity and linear distance from the horizon.  From \ref{c=slogs}  and \ref{cstar} we see that the Planck distance is identified with the layer on which  complexity equals the scrambling complexity $\c_{\ast}.$

 An obvious question is what happens when complexity exceeds the scrambling value? Does the distance-complexity relation break down or is there a meaning to sub-planckian distances? For the moment I will just finesse the question by defining the distance from the horizon in terms of the degree of complexity; in other words by \ref{fundformula}.
Later we will come  back to the  the meaning of sub-planckian distances.

\subsection{Gravity and Complexity}

Entropic theories of gravity \cite{Jacobson:1995ab}\cite{Verlinde:2010hp}  build in various ways on the parallels between general relativity and thermodynamics, and on entropic forces in statistical mechanics. Undoubtedly there is  truth to these ideas. I want to suggest that there may be another deep connection; this time between gravity and complexity. To state it as a slogan:

\bigskip

\it

Things fall because there is a tendency toward complexity.
\rm

\bigskip

This sounds  like the   philosopher who said that ``all bodies move toward their  natural place,"  but I think it's better.
\bigskip

Let's begin with the fact that complexity satisfies a law similar to the second law of thermodynamics. It tends to increase, but with caveats that are similar to those that apply to the increase in entropy. A particular example goes as follows: Consider  a simple operator $W$ of low complexity such as  a single qubit operator, or an undecorated  single-plaquette Wilson-loop. Suppose we evolve $W,$

\be
W(t) = U(t) W U^{\dag}(t).
\ee

\bn
One  expects the complexity to increases whether we evolve forward or backward in time. The symmetry follows from the fact that an undecorated Wilson loop is time-reversal invariant.

Now consider that the complexity is a monotonic function of distance from the horizon; being largest very near the horizon, and decreasing toward the boundary. We can express the trajectory of complexity as a real trajectory in space. The excitation described by acting with $W$ begins in the  past, moving away from the horizon (complexity decreasing). At $t=0 $ the excitation reaches a maximum distance from the horizon (minimum complexity), and then falls back toward the horizon (increasing complexity). This of course is exactly what we expect for a particle moving in a gravitational field.

Aristotle's theory of gravity was first-order in derivatives; \it force is equal to mass times velocity. \rm  \ The complexity-based theory looks more like a second order theory because complexity and it's rate of change are independent variables. Take an operator with greater complexity than the minimum, but choose it in a time-reversal invariant way. An example is a large undecorated Wilson loop. The complexity of such an operator is stationary. For example if it acts at
 $t=0$ the complexity will satisfy
  \be
  \c(t) = \c(-t).
  \ee

  \bn
  Thus the complexity can be  large, but  have zero time-derivative. If evolved forward or backward the  complexity will increase.

If the complexity   is not stationary it will either increase or decrease. If it increases it will continue to increase, but if it decreases that will not last long. It will reach a minimum and turn around.   The same thing is true in a time-reversed sense if the complexity is increasing at $t=0.$ If we run it back in time it will also reach a minimum and turn around.

The trajectories all follow the same overall tendency of rising (in the sense of moving away from the horizon) to a minimum and then falling toward increasing complexity. And even if an operator begins with decreasing complexity it will quickly turn around and fall toward increasing complexity.

\sc
\section{The Two-Sided  Case}

\subsection{Why Start with the Thermofield-Double State?}

When a real black hole forms, it does not begin its life in a random state. Initially it is far from equilibrium, and its horizon and interior are perfectly vacuum-like. If our goal is to understand how these objects evolve then whatever model we employ  should preserve these  features.
The first feature is the initial smooth horizon.  This is important because there is no known mechanism for a smooth horizon to become singular, i.e., for a firewall to form.

The second feature requires some explanation.
Maldacena has suggested\footnote{J. Maldacena, private communication.} that the stability of smooth horizons may be analogous to stability of cosmological space-times. Cosmological solutions tend to be stable while they are expanding and unstable when collapsing.   Maldacena draws an analogy with black holes in which the role of cosmological expansion  is played by the stretching, as a function of time, of certain spacelike slices in the black hole interior.

\begin{figure}[h!]
\begin{center}
\includegraphics[scale=.3]{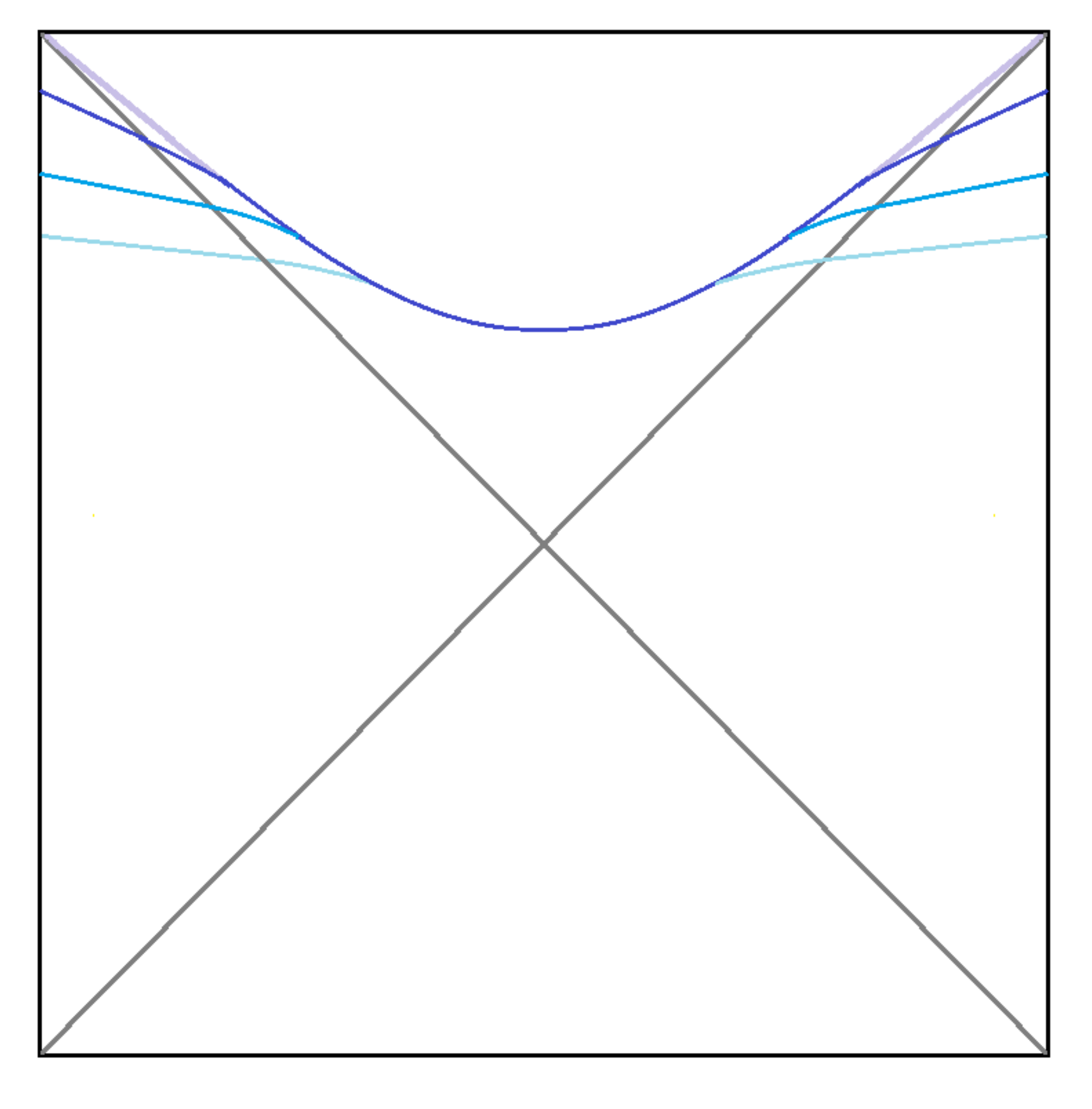}
\caption{As time evolves symmetrically in the eternal black hole, the ERB is stretched in length. The stretching takes place near the horizons. }
\label{stretch}
\end{center}
\end{figure}

The stretching  phenomenon was discussed in \cite{Hartman:2013qma}\cite{Qi:2013caa} and is illustrated in figure \ref{stretch} (see also figures 2 and 12 of reference \cite{Hartman:2013qma} ).    Each surface can be labeled by the time at which it intersects the boundary. The stretching takes place near where the blue spacelike slices intersect the horizon and the length of the surface $L(t)$ grows linearly with the time,

\be
L(t) \sim l_{ads} t.
\ee

\begin{figure}[h!]
\begin{center}
\includegraphics[scale=.5]{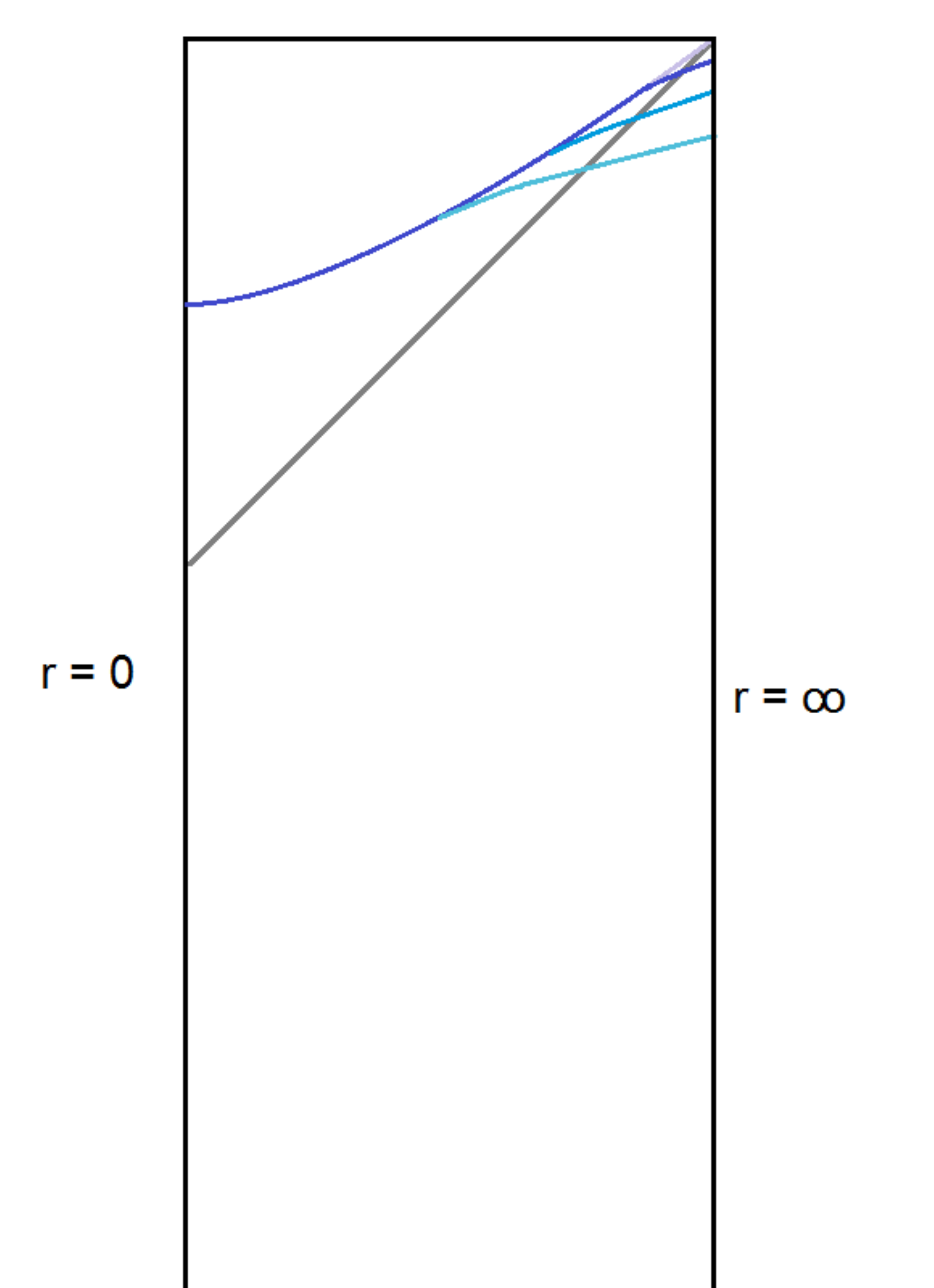}
\caption{Although there is no second side for a one-sided black hole, we may think of the interior  a bridge. Stretching of the same kind as in figure \ref{stretch} takes place for a one-sided black hole created by collapse.  }
\label{onestretch}
\end{center}
\end{figure}

\bn
Stretching is not restricted to  two-sided eternal black holes. It also takes place in one-sided black holes  as illustrated in figure \ref{onestretch}. Stretching has the effect  of  red-shifting and diluting perturbations close to the horizon in the interior of the black hole.

The rate of stretching in a real black hole is positive, but in ADS black holes it must be the case that contracting is also possible. The time-reversal invariance of the dual  CFT insures that contracting states are as abundant as stretching states.

Given the importance  that initial  stretching plays  in stabilizing the horizon it should be preserved in any attempt to model realistic black holes\footnote{The demand that the interior has positive stretching at the formation of the black hole is closely related
 to a censorship rule recently proposed by Page \cite{Page:2013mqa}}.

This leads to the  following conjecture:

\bigskip

\it

\bn Black holes that start with smooth horizons, and with  interior stretching, will stay smooth.

\rm

\bn
We may have to qualify this for ADS black holes by adding the phrase \it at least for an exponentially long recurrence time.\rm \ The reason will be discussed shortly.

Consider the usual eternal ADS black hole. The full Penrose diagram does not satisfy the stretching criterion, since the lower half of the diagram is contracting. For that reason we will consider the lower half to be fictitious. We
 assume that the two-sided  system was created on a time-slice in the future half of the geometry. The most dangerous situation from the firewall perspective would be to start the  history on the $t=0$ slice in the Thermofield-double state (In bulk terms, the Hartle-Hawking state). At that time the stretching rate is zero, and the left and right bulk geometries are as close as possible \cite{VanRaamsdonk:2010pw}.

 We are interested in what Bob sees when he jumps into his black hole.
 Given that the black holes are created at $t=0$ Bob's jump-off time must be
 positive. Likewise, Alice will be restricted to acting on her side of the system  at positive time. However, she can  simulate acting at earlier times  by acting with precursors \cite{Susskind:2013lpa}.

 Shenker and Stanford have identified a wide class of states that have smooth horizons and expanding ERB's \cite{Shenker:2013yza}. Most of them are better protected from accidental or intentional firewall production than the TFD because their Penrose diagrams are wider than that of the TFD  \cite{VanRaamsdonk:2010pw}. This makes it harder to send a signal through the ERB. It seems that the most favorable entangled state for sending firewalls,  which satisfies the two criteria described above, is the TFD.

\subsection{Dual Description of Stretching}

\begin{figure}[h!]
\begin{center}
\includegraphics[scale=.3]{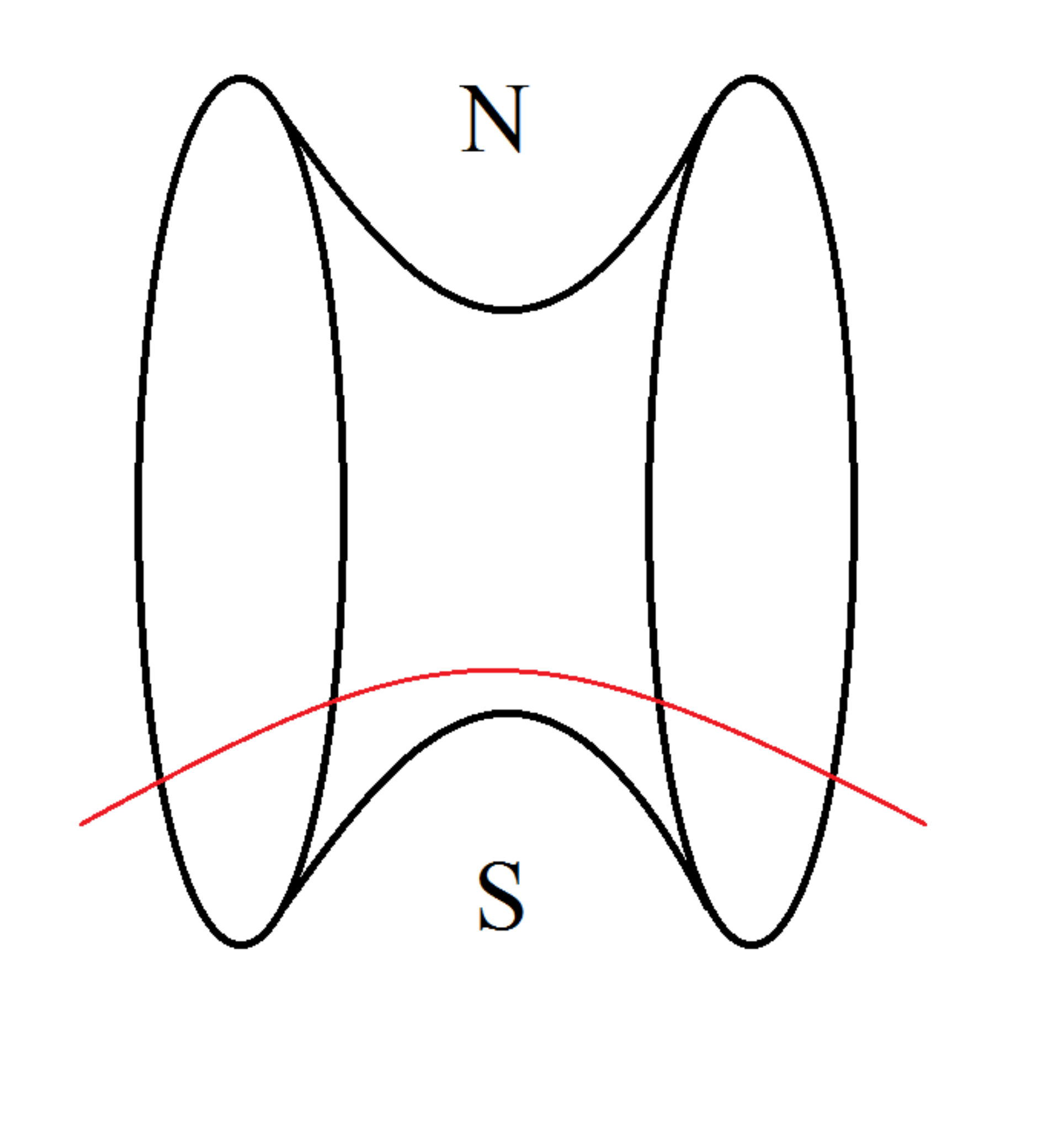}
\caption{The spatial geometry of the TFD is two asymptotic regions connected by an ERB. The left and right sides are maximally entangled but there is very little vertical entanglement across the red line.   }
\label{vertical}
\end{center}
\end{figure}

The stretching of an ERB is easy to describe in the bulk description of  ADS black holes, but it is not  clear
what it means in the dual gauge theory description.  In \cite{Hartman:2013qma} the authors relate the stretching of ERB's to the growth of ``vertical" entanglement  in a two-sided ADS black hole \cite{Maldacena:2013xja}. Vertical entanglement refers to  figure 9 of \cite{Maldacena:2013xja} which is reprinted here as figure \ref{vertical}.
The TFD state is maximally entangled horizontally it has very little vertical  entanglement between different angular regions (see figure \ref{vertical}).

One might try to use the growth of vertical entanglement as a diagnostic for stretching but there is a problem with this. The vertical entanglement becomes maximal very quickly. By the scrambling time it has reached its maximum value\footnote{Douglas Stanford pointed out that the order $N^2$ part of the vertical entanglement saturates in a time of order $l_{ads}$ which is even smaller than the scrambling time. }, but the ERB continues to grow. Classically the ERB  stretches forever.

What we are looking for is a gauge theory quantity to represent the length of the ERB. It should grow for as long as we trust the classical evolution; and it should be defined for both  the one-sided and two-sided cases. Entanglement  is too crude a measure of complexity. It reaches its maximum at the scrambling complexity, which as we have seen, is very far from the maximal complexity that can be attained.

A quantity that has the right properties  is the computational
complexity per qubit  of the state $$\frac{\c}{S}, $$  either  two-sided or one-sided.
Thus far I have discussed the complexity of unitary operators, but it is also possible to define the complexity of states. To do so we have to begin with a concept of a simple state. For a one-sided system of qubits the definition of a simple state is a product state in the computational basis. For example the state $|000000....\ra$ may be regarded as simple. It is a product state with no entanglement, and it is easy to construct.

For a two sided system with maximal horizontal entanglement, the simplest state is one with no vertical entanglement. This means a product of $K$ Bell pairs, each shared between the two sides. For example,

\be
|\psi\ra = \{ |00\ra + |11\ra \}^{\otimes K}
\label{simple}
\ee

This is a simple qubit analog of the TFD. It is not as simple as the one-sided state, but it is also easy to make if one has a source of Bell pairs.

Now consider a more  general state. In the one-sided case it can be any quantum state: in the two sided case it can be any quantum state that can be prepared by local operations\footnote{Local operations means unitary transformations which have the form of a product of a left-side unitary and a right-side unitary.} on \ref{simple}. Consider generating these states by feeding the simple states into a quantum computer and allowing gates to act on the states. The obvious definition of complexity of a state is the minimal number of gates  needed to prepare the state.

The complexity, defined in this way, continues to grow long past the scrambling time, but not forever. Eventually at the classical recurrence time it reaches its maximum. There is independent reason be believe that the classical geometry becomes unreliable at the recurrence time \footnote{D. Stanford, private communication.}.

Once the recurrence time has been reached the complexity fluctuates, sometimes increasing and sometimes decreasing. On very long (doubly exponential) time scales it can return to small values. If we assume that the rate of change of complexity is the dual description of stretching, then over multiple recurrence times the ERB will stretch and shrink. During the shrinking periods the horizon is vulnerable to the formation of firewalls unless the state is fine-tuned to be very close to  the TFD.

For most of the rest of this paper the time scales will be assumed to be much shorter than the classical recurrence time. If we assume that the black hole starts in a simple state the complexity will increase for all polynomial time scales.

\sc
\section{ Sending Firewalls is Hard}

Black holes with non-smooth horizons are definitely possible. Suppose we try to create a pair of entangled black holes in the TFD, by going back to a large negative time and preparing the  system in the state,

\be
| preTDF  \ra = e^{iHt} |TDF\ra
\ee

\bn
where $|preTDF\ra$  is by definition a state that will evolve to the $TFD$ in a time $t,$ and $H=H_L +H_R$ is the total Hamiltonian of the right and left systems. As I explained, this requires a great deal of fine-tuning; a mistake on one qubit will cause the system to veer off to a completely different state---even a completely different classical geometry. Some mistakes will lead to a firewall for Bob's black hole, at least temporarily.

Building on the work of Von Raamsdonk \cite{VanRaamsdonk:2013sza}, Shenker and Stanford \cite{Shenker:2013yza} have  described how minor perturbations on Alice's side (the left)  at large negative time can create high-energy shockwaves  which propagate to Bob's side (the right). If Bob jumps through the horizon he  will encounter a firewall. This means that arguments which say that firewalls never occur must be wrong. Curiously Shenker and Stanford find that if early  perturbations take place on both sides, the effect is to smooth the horizon \cite{Shenker:2013yza}.

The question we will ask is this:

\it
\bn
Given that the two-sided system starts in the TFD, how hard is it for Alice to create a high-energy shockwave  behind Bob's horizon?

\rm

\bn
For the moment let us ignore the fact that the past history of the TDF is fictitious and consider the entire extended Kruskal diagram to be physical, keeping in mind that it is a useful fiction.  Alice may act on the left side at any time, past or future.

\begin{figure}[h!]
\begin{center}
\includegraphics[scale=.3]{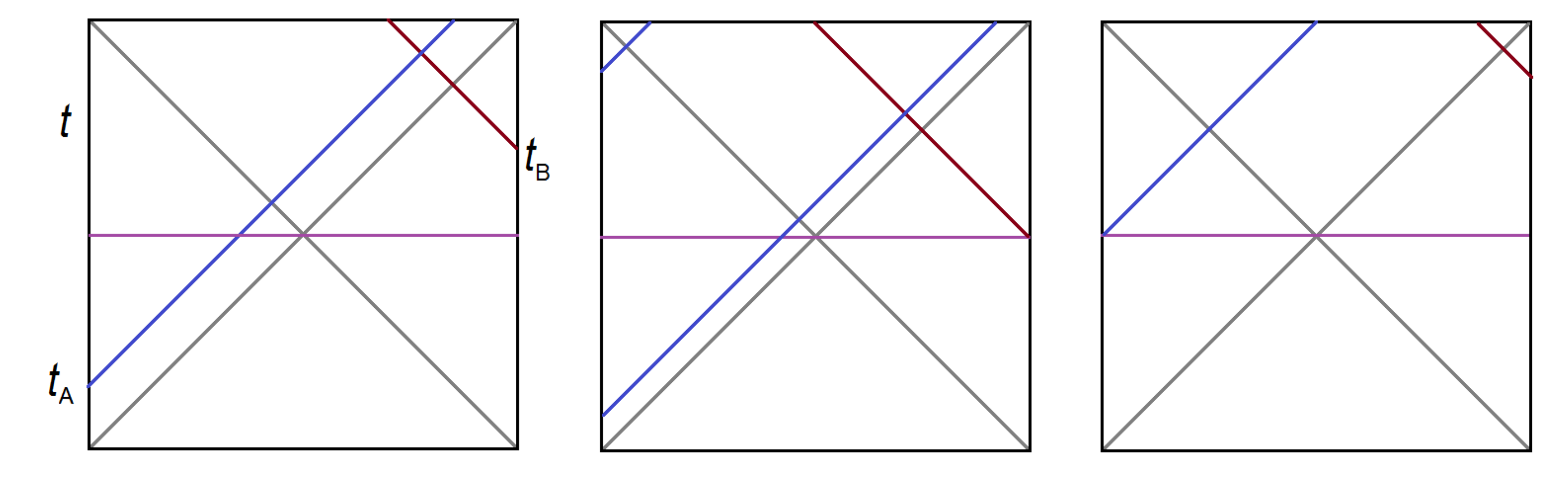}
\caption{ In the first panel Bob jumps in at $t_B$ and Alice sends a signal from $t_A.$  In the center panel the figure has been boosted so that $t_b=0.$ In the final panel it has been boosted so that $t_a=0.$}
\label{e81}
\end{center}
\end{figure}

The setup we will consider is illustrated in the first of figures \ref{e81}.
There are three times labeled on the figure.
They represent:

\bigskip

\bi

\item The  time  $t_A$ at which Alice acts with a simple operator $W$ of low complexity. If $t_A<0$ then it should be regarded as fictitious.  In the Shenker-Stanford context $t_A$ is the time that Alice launches  a perturbation $W.$ The launch time may be negative, zero, or positive.

\item The time at which Alice acts with a precursor \cite{Susskind:2013lpa} called $t.$  $t$ is always positive or zero. The precursor has the form

    \be
W_p = U(t - t_A) \ W \  U^{\dag}(t - t_A).
\label{Ep}
\ee

One may say that if $t_A<0$ then acting at time $t$ with \ref{Ep} ``precurses" the  action of $W.$

    \item The time  $t_B$  that Bob jumps into the black hole from the boundary. We take $t_B$ to be positive or zero.

\ei

\bn
For the moment we can ignore $t$ and concentrate on $t_A$ and $t_B.$ The symmetry of the $TFD$ implies that the effect of Alice's action on Bob  depends only on the combination $t_A + t_B.$ For that reason we don't lose any generality by setting $t_B$ to zero.

 The operator $W$ is
not strictly local;  a local operator on the boundary of ADS carries infinite energy. The perturbations considered in  \cite{Shenker:2013pqa} have energy of order the black hole temperature. To create a low energy excitation a local boundary operator must be integrated over time as in \ref{smear}. It will have a complexity of order unity, and be  localized in the bulk at distance $\sim l_{ads}$ from the horizon.

If Alice acts at $t_A> 0,$ and if Bob jumps at   $t_B=0$ from his boundary,  then an examination of the usual Penrose diagram shows that Alice's signal cannot reach Bob before he hits the singularity. This is true, at least for timescales over which we expect the semi-classical geometry of the Penrose diagram to make sense\footnote{How long this is is an open question. There are  indications that the geometric description must break down by the classical recurrence time $e^S.$}.
Thus Alice must act at negative $t_A$ to get the signal to Bob. Let us therefore take $t_A<0.$

For $t_A$ in the range $$-t_{\ast} < t_A  <0,$$ the Shenker-Stanford analysis is reliable but the effect on Bob is mild. However, the energy of the signal increases as $t_A$ becomes increasingly negative. As $|t_A| $ approaches the scrambling time the signal becomes progressively more energetic and by $t_A = -t_{\ast} $ it has achieved the status of a Planckian shockwave. Beyond that the quantum corrections grow and become out of control.

One may think that this trend continues for arbitrarily negative time. However that cannot be the case. If one goes back to sufficiently negative $t_A$  the quantum recurrence theorem implies that the effect will be the same as acting at slightly positive $t_A,$ for which Bob experiences  nothing.

One other point  to note is that the effect of the shockwave at Bob's end is temporary. It is easy to see that if Alice disturbs the system at time $-t_A,$ then the shockwave will have disappeared at Bob's end by time $+t_A.$ What happens to it is that it falls into the singularity.

\subsection{Why is it Hard?}

We now come to the central point: why it is hard for Alice to send a destructive Planckian shockwave to Bob. The most dangerous situation for sending such signals, subject to the constraint that of positive stretching, is to assume the system was  created in the TFD at $t=0$ and allow Alice to act with the precursor immediately after. The longer Alice waits the harder it will be to send a shockwave.

In considering how difficult it is for Alice to send a signal we must take into consideration that the past is fictitious and that Alice must send the signal at time $t>0$ by means of a precursor. For the signal to be a Planckian shockwave by the time it reaches Bob  it must have been launched at a time $t_A \sim -t_{\ast}$
or earlier. This means that the precursor must have had complexity $\sim \c_{\ast} + t/l_{ads}$ which is beyond the scrambling complexity. Although very small by comparison with the maximum complexity, the scrambling complexity is still very complex. To get a rough idea of what it means, let's consider  a  perturbation of the Hawking radiation which perturbs one qubit of Alice's stretched horizon. This corresponds to the action of a unitary operator with complexity $=1.$ The same amount of  complexity applied to the formula  \ref{fundformula}
would give a distance $l_a$ from the horizon satisfying,

$$\frac{l_a^2}{l_{ads^2}} = e^{-\c / S}= 1$$

\bn
such a perturbation is located in the bulk at a macroscopic distance $\l_{ads}$ from the horizon.

Next consider a generic perturbation of all $S$ qubits. An example of a generic perturbation would be a product of $S$ qubit operators. A perturbation of this type  will have complexity $S.$ Equation \ref{fundformula} gives

\be
\frac{l_a^2}{l_{ads^2}} = e^{-\c / S}= 1/e.
\label{uncomplex}
\ee

\bn
or

\be
l_a = .6 \ l_{ads}.
\label{la=.61 lads}
\ee

\bn
By going from complexity $1$ to complexity $S$ we have penetrated into the stretched horizon by only a factor of $.6.$

On the other hand complexity $S \log S$ brings us all the ways to the Planck distance, and even modestly larger complexity would bring us into the realm of trans-scrambling complexity.

\begin{figure}[h!]
\begin{center}
\includegraphics[scale=.3]{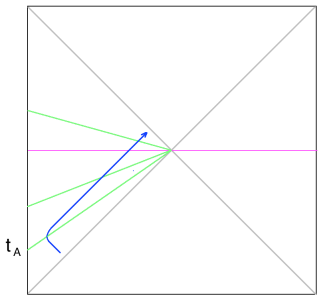}
\caption{ By means of a precursor, Alice acts at $t_A$ with a minimally complex perturbation. Because the complexity increases with time, the signal moves toward Bob.}
\label{success}
\end{center}
\end{figure}

Let us consider in a little more detail, the process of sending a specific message to Bob---`` Happy Birthday Bob."
Let's fix the time $t_A$ in such a way that the signal reaches Bob's side with some particular energy, and then allow $t$ in equation \ref{Ep} to vary. At $t_A$ (the fictitious launch time) the signal was minimally complex and localized at the edge of the stretched horizon (see figure \ref{success}. Clearly the complexity of the precursor  $W_p$ increases with $t$ from that point, implying  that the signal moves inward toward the horizon.    The increasing complexity translates into motion \it toward \rm  \ the horizon and toward Bob.
So as expected, Alice succeeds in sending a signal through the ERB which Bob can receive if he crosses the horizon. Note that it is not just the degree of complexity that matters. It is the \it direction in which the  complexity changes, \rm  \ which determines whether the signal propagates toward or away from Alice's horizon.

Now consider the degree of difficulty of applying $W_p.$  This can be confusing because $W_p$ is defined so as to precurse the effect of a past action. It may be helpful to time-reverse Alice's problem (Turn figure \ref{success} upside down). In the time-reversed version Alice must act with a complex operator at a certain time, and thereby  produce the same effect as a simple operator at a later time. In general, without fine-tuning, complexity increases with time. So it is evident that Alice must fine-tune the precursor if she wishes to produce the same effect as the simple operator $W$ at a significantly later time. The larger the time difference the greater the degree of fine-tuning.

We can turn the problem back over and ask how much fine-tuning is required for Alice to precurse the effect of a simple operator at an earlier time. The answer is exactly the same. The larger the separation between $t$ and $t_A,$ the greater the required fine-tuning. If, after waiting a scrambling time  ($t > t_{\ast}$) Alice wishes to send a signal to Bob, she must apply extraordinary care in sending the message. An accidental mistake on one qubit out of $S$ will ruin the message.

We can quantify this  in a little more detail by supposing that there is a probability $\epsilon$ for an error in each gate (See appendix A for the meaning of an error). The probability that no error occurs in Alice's attempt to send the message is

\be
P = e^{-\epsilon \c}
\ee

\bn
where the complexity $\c$ is the number of gates that Alice's quantum computer has to apply. At the scrambling time

 $$\c = S \log{S}.$$

\bn
For a solar mass black hole $S \sim 10^{76}$ so that

\be
\epsilon < 10^{-78}.
\ee

\bn
in order for the probability of Alice's Birthday message getting through to be appreciable.
Things get worse  with the time separation between $t$ and $t_A.$

These estimates were for Alice to be able to send a specific message to Bob. But to send Bob a nasty shock does not require
the signal to contain a recognizable message. Errors may ruin the message but allow a blast of energy to hit Bob. Thus it is important to understand how errors affect the signal\footnote{I would like to thank an anonymous Berkeley student for clarifying the difference between sending a specific message and sending a shockwave. }.

\begin{figure}[h!]
\begin{center}
\includegraphics[scale=.3]{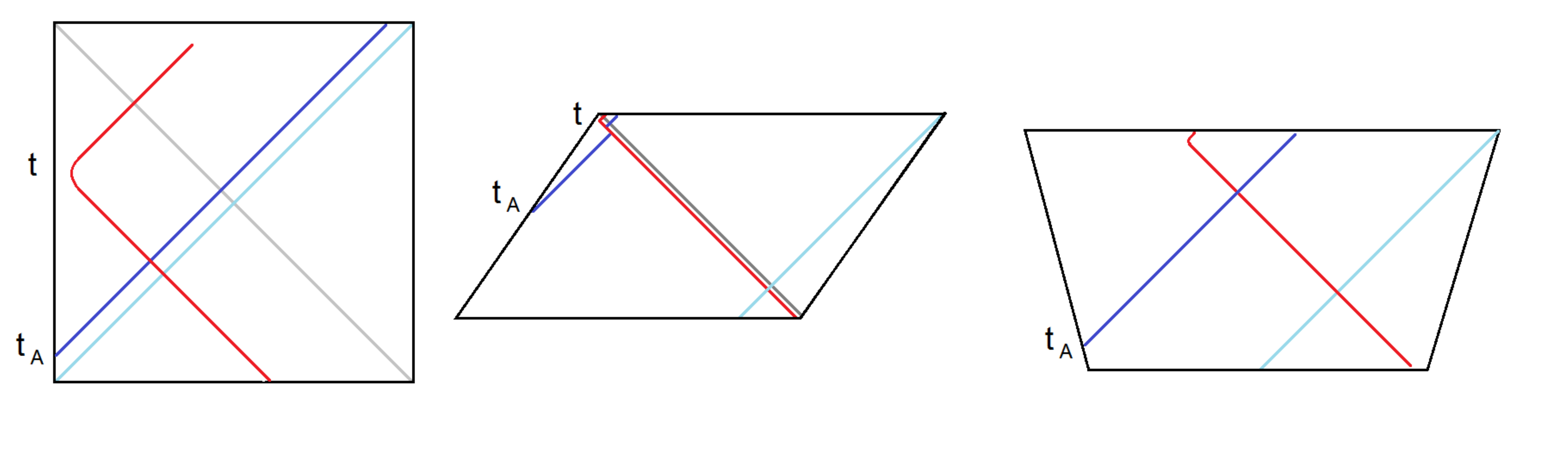}
\caption{ The effect of an error on Alice's attempt to signal Bob is the addition of a second Shenker-Stanford shock shown in red. The first diagram is accurate if $t_A$ is not too negative. But as $t_A$ becomes more negative the collision becomes higher energy and creates a back-reaction on the geometry. The second diagram shows the effect of the error-shock wave. The diagram is not a true Penrose diagram. Light rays such as the blue ray are shifted as they cross the red shockwave. A true Penrose diagram accounting for both shock waves would look more like the last diagram (same as figure 2 of \cite{Shenker:2013yza} ). The effect of the back reaction due to the error is to decrease the energy of the shockwave in Bob's frame. }
\label{shift}
\end{center}
\end{figure}

We can model an error  by multiplying the precursor $W_p$  by a one-qubit operator that I will call $e$ for error.
The order that $W_p$ and $e$ are applied is important. For $e$ to affect the signal we must apply the operators in the order $W_p \ e.$
The effect is the same as applying the product of $W$ at time $t_A,$ and $e$ at time $t,$ \it except \rm that the operators are not time-ordered. This is exactly the type of problem that Shenker and Stanford studied in \cite{Shenker:2013yza}. According to their analysis the result of the error is a second shockwave that collides with the message in a way that is naively summarized by figure \ref{shift} (See also figure 2 of \cite{Shenker:2013yza} ).

The diagrams show that the collision results in a delay of the message and an increase in the distance between the signal and Bob's horizon. This increased separation implies a decrease in the energy of the signal when it reaches Bob. Errors clearly diminish the effect of the signal. Thus Alice must fine-tune her signal and avoid too many errors even if she is just trying to send a blast of particles with no message.

To compensate for the possibility of errors  Alice could make $t_A$ more negative. But this would be at the cost of more complexity.

\bigskip

\begin{figure}[h!]
\begin{center}
\includegraphics[scale=.3]{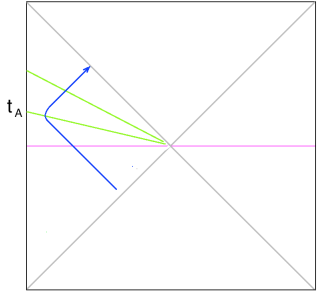}
\caption{ Alice acts with the simple operator $W$ at time $t_A >0.$ This time the complexity would decrease away from $t=0,$ so the signal initially would propagate away from Bob.}
\label{fail}
\end{center}
\end{figure}

Now let's turn to  the case in which Alice acts with a simple operator $W$ at positive time (see figure \ref{fail}).
This time Alice's action at $t_A$ is not fictitious. There is no particular fine tuning; Alice applies the simple operator $W.$ However we may ask if there
is an operator that a fictitious Alice could act with at an earlier time, say  $t=0,$ which would have the same effect as acting with $W$ at $t_A.$. The answer is of course the precursor,

\be
W_p =  U^{\dag}(t_A) \ W \ U^{\dag}  U(t_A )
\label{Ep reverse}
\ee

\bn
This time the earlier operator at $t=0$ is complex and the later operator at $t_A$ is simple. Geometrically the signal propagates from near the horizon (complex) toward the boundary (simple), and therefore away from Bob. Of course it will bounce back at $t$ but by that time it is too late to reach Bob.

To summarize, sending a signal to Bob from a positive time requires a high degree of complexity and an operation which is fine-tuned in order to reverse the arrow of time at Alice's black hole. The later the signal is sent, the greater the degree  of complexity and fine-tuning.  On the other hand acting with a simple operator at positive time requires no fine tuning,  but it also does not send a signal to Bob.

\subsection{Comment on Simple and Complex Disturbances}

The difference between disturbances which can and cannot send signals through an ERB was emphasized in \cite{Maldacena:2013xja}. The disturbances which cannot send signals were called simple and those that can were called complex\footnote{The terms easy and hard were used in \cite{Susskind:2013aaa}.} but no clear criteria for the difference was proposed but it was suggested that complex operators were anything that involved more than half Alice's qubits. According to this viewpoint, by measuring all Alice's qubits, a firewall could be created at Bob's end.

In this paper a sharp criterion based on complexity is being proposed. What does it imply if Alice measures all the qubits on her side? The answer is that it has very little effect on Bob's side. As we saw earlier in \ref{uncomplex} and \ref{la=.61 lads} an operator that involves $S$ qubits in a simple product form is not significantly more complex than a single qubit operator. Thus as difficult as it seems, measuring all $S$ qubits should be classified as a simple or easy disturbance.

\sc
\section{Trans-scrambling Complexity and Generic Black Holes}

\subsection{Sub-planckian distances}

In the frame in which Bob jumps at time $t=0$ a shock wave may be characterized by its complexity. The complexity can be anything from negligibly small to the maximum $\c_{max} = e^S.$
According to formula \ref{fundformula}  distances $l_a$ as small as

$$
\exp{-e^{e^S}}
$$

\bn
make kind of sense. For most of this vast sub-planckian range of scales we don't expect that the operational meaning has anything to do with meter sticks. Let us consider what meaning they do have. It has more to do with large times than small distances.

\begin{figure}[h!]
\begin{center}
\includegraphics[scale=.3]{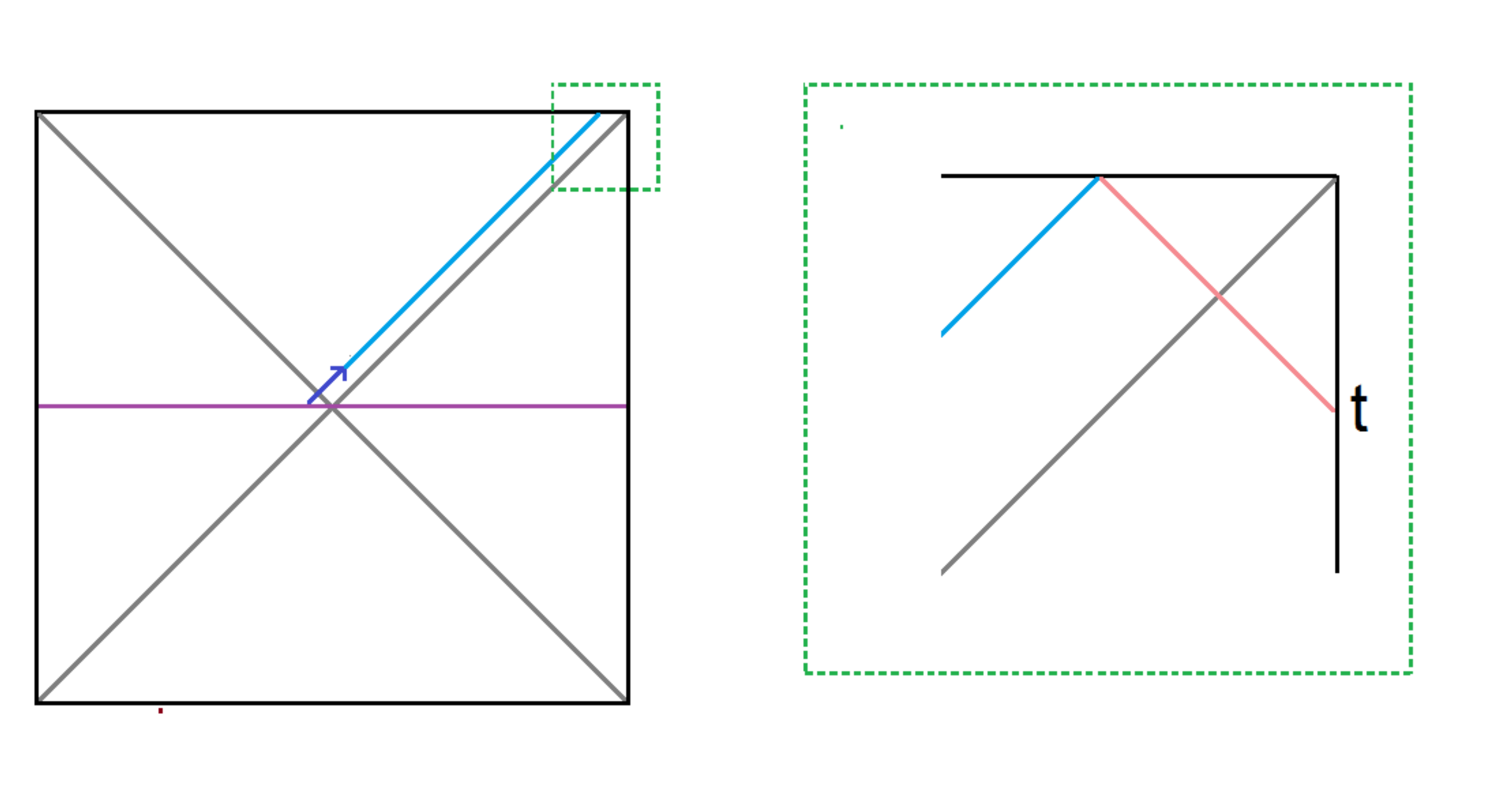}
\caption{Shock wave created by Alice falls into the Singularity. The last moment that Bob can experience the shockwave is time $t.$}
\label{e12}
\end{center}
\end{figure}

To keep things simple let's choose the time that Alice acts with the precursor $W_p$ to be $t=0.$ The complexity of the precursor
determines a location on the $t=0$ surface according to \ref{fundformula}. We can also characterize the precursor by the time at which the fictional simple operator $W$ acts. Let's call it $-t.$ Assume that $t<< e^S.$ The shockwave created by $W$ is shown in figure \ref{e12} originating on the $t=0$ surface. The shockwave eventually falls into the singularity at a time determined by the complexity of the precursor $W_p.$

The operational meaning of the distance $l_a$ is that it determines the length of time that the shockwave or firewall persists near Bob's horizon. This connection breaks down when the complexity reaches its maximum. From Bob's point of view that time is the classical recurrence time. For smaller times the lifetime of the shockwave is given by

\be
t = \frac{\c}{S} l_{ads}.
\label{lifetime}
\ee

\subsection{Generic Black Holes}

For the rest of this section we  consider one-sided ADS black holes formed by sending in energy from the ADS boundary.
It is well known that the entire set of $e^S$ states of a black hole cannot be made by rapid collapse. The states made quickly are not generic, and form a small subspace. It is possible that black holes which form by collapse have smooth horizons,
even if  generic black holes have firewalls.

The following argument has been made:

\bn
The entire space of $e^S$ states can be accessed by building the black hole slowly, one quantum at a time. The idea is to construct the black hole by a process which roughly time-reverses evaporation. To build a generic state of a Hawking Page black hole in this way would take a time of order $\l_{ads} \ S.$ The argument sais tht if generic states have firewalls then the states formed this way have firewalls.

Here is the counter argument based on complexity. It can be shown \cite{Knill} that almost all states have exponentially large complexity. This means that they cannot be formed with fewer than $e^S$ gates. To put it another way, almost all states can only be formed exponentially slowly. The class of states that can be made in time $\l_{ads}$ is still very special and not representative of generic states. Even if generic states have firewalls that does not mean the states formed over a time $l_{ads} \  S$ do.
One should also note that for ADS black holes  the complexity stops increasing at the classical recurrence time, implying that the ERB stops stretching. This makes the horizon more vulnerable than during the time
that it stretches. Conversely, before the recurrence time there is plenty of room for the complexity to increase so that stretching can protect the horizon.

There does not seem to be any reason that black holes that are formed over times much less than the recurrence time should have 
firewalls.

\sc
\section{Continuation to Interior}%%%%%%%%%%%%%%%%%%%%%%%

Representing bulk physics by boundary data at a fixed time is difficult for well known reasons. It is especially difficult to represent the bulk degrees of freedom behind the horizon. And finally it is extraordinarily difficult to represent them at late time. The difficulty is connected with complexity and chaos. The purpose of this section is to explain these difficulties.

\begin{figure}[h!]
\begin{center}
\includegraphics[scale=.3]{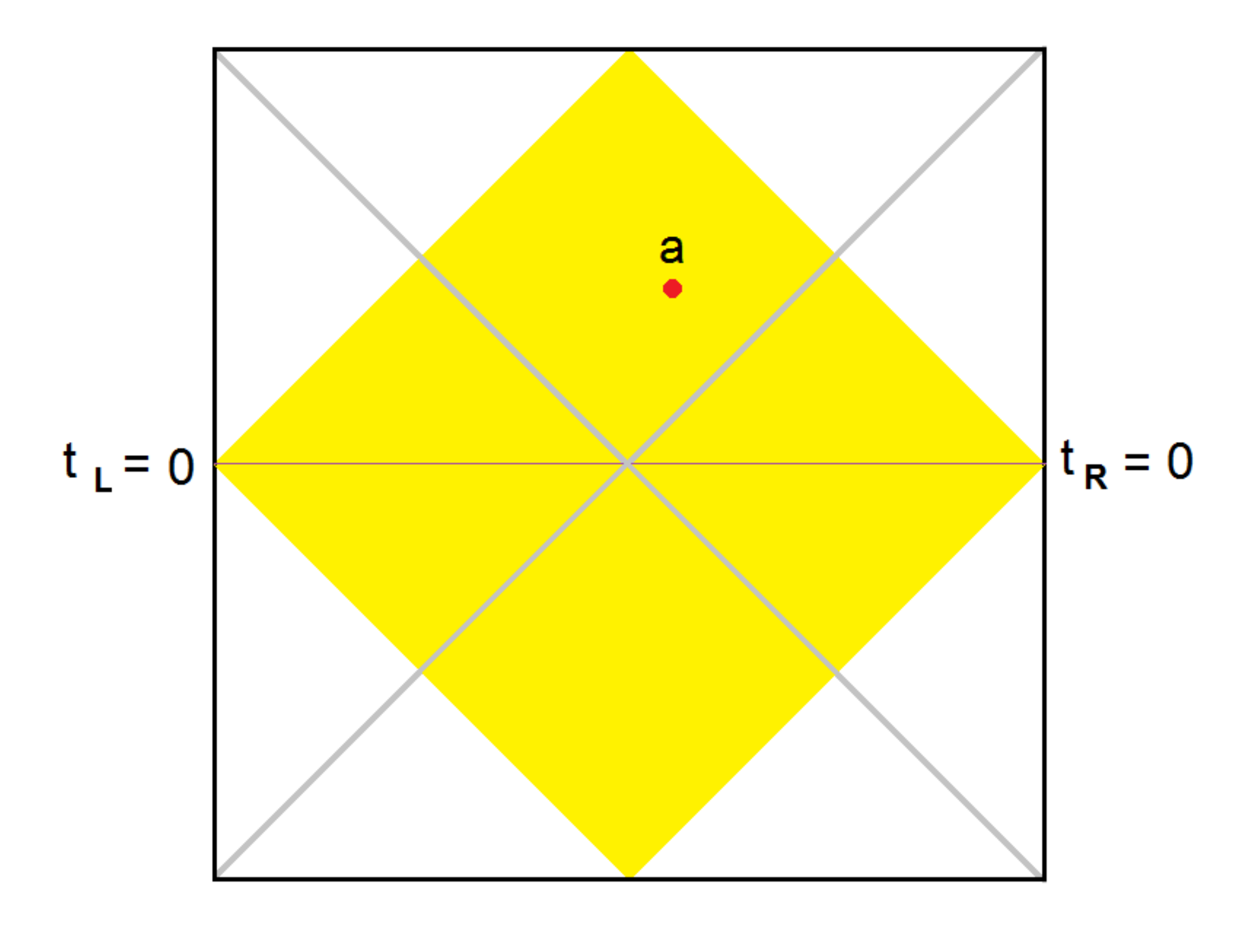}
\caption{The yellow diamond is the Wheeler-DeWitt patch represented by the two-sided CFT system at $t=0.$ The red dot is the point ${\bf{a}}$ behind both horizons.}
\label{wdw1}
\end{center}
\end{figure}

To begin with we consider the meaning of the instantaneous state-vector
 of the two decoupled (but entangled) CFT's. The CFT state does not encode a bulk state  vector on any particular slice. What it does represent is  an entire Wheeler DeWitt history \cite{Maldacena:2001kr}\cite{Maldacena:2013xja}. The history comprises all spatial slices that end at a given pair of times $t_L$ and $t_R$ on the two boundaries. In figure \ref{wdw1} the WDW patch (yellow region) corresponding to the times $t_L = t_R = 0$ is shown along with a point ${\bf{a}}$ (in red).

Our first goal is to to express the
 low energy bulk field $\phi(a)$   in terms of gauge theory operators at $t_L = t_R = 0.$ For the moment the point  ${\bf{a}}$ is chosen to be located in a place that avoids certain extremes. We don't want ${\bf{a}}$ to be too close to the bifurcate horizon: we will   assume it is many planck lengths away. We also will keep $\phi(a)$ far from the singularity. Finally we require that $\phi(a)$  is not far out, near the upper corners of the Penrose diagram. We can express this condition in terms of the hyperbolic angle that ${\bf{a}}$  is displaced from the vertical axis: the boost angle should be much less than $\log{S}.$

\begin{figure}[h!]
\begin{center}
\includegraphics[scale=.3]{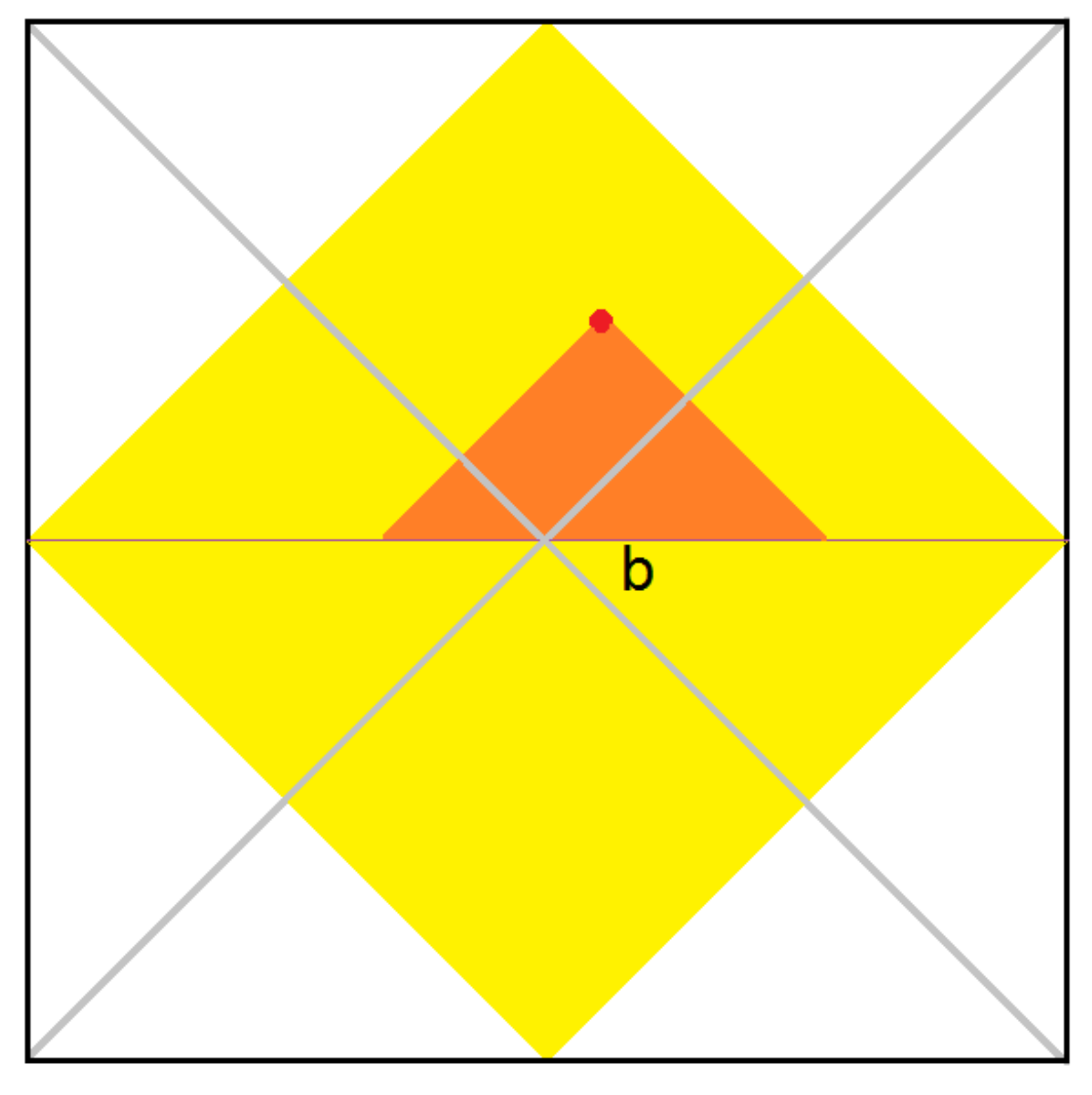}
\caption{The causal past of point  ${\bf{a}}$ is shown in orange. It intersects $t=0$ on the base of the triangle $b.$}
\label{wdw2}
\end{center}
\end{figure}

 The construction begins with a pull-back operation. The low energy bulk field equations are used to express $\phi(a)$ in terms of the field on the $t=0$ surface. Figure \ref{wdw2} shows the causal past  of ${\bf{a}}$ (orange) and its intersection with the $t=0$ surface $b.$  The field $\phi(a)$ may be expressed as an integral of fields on  $b.$
\be
\phi(a)  = \int_{b} G(b,a) \ \phi(b)
\label{int-rep}
\ee

 \bn
 where $G(b,a)$ is some Green's function.

\begin{figure}[h!]
\begin{center}
\includegraphics[scale=.3]{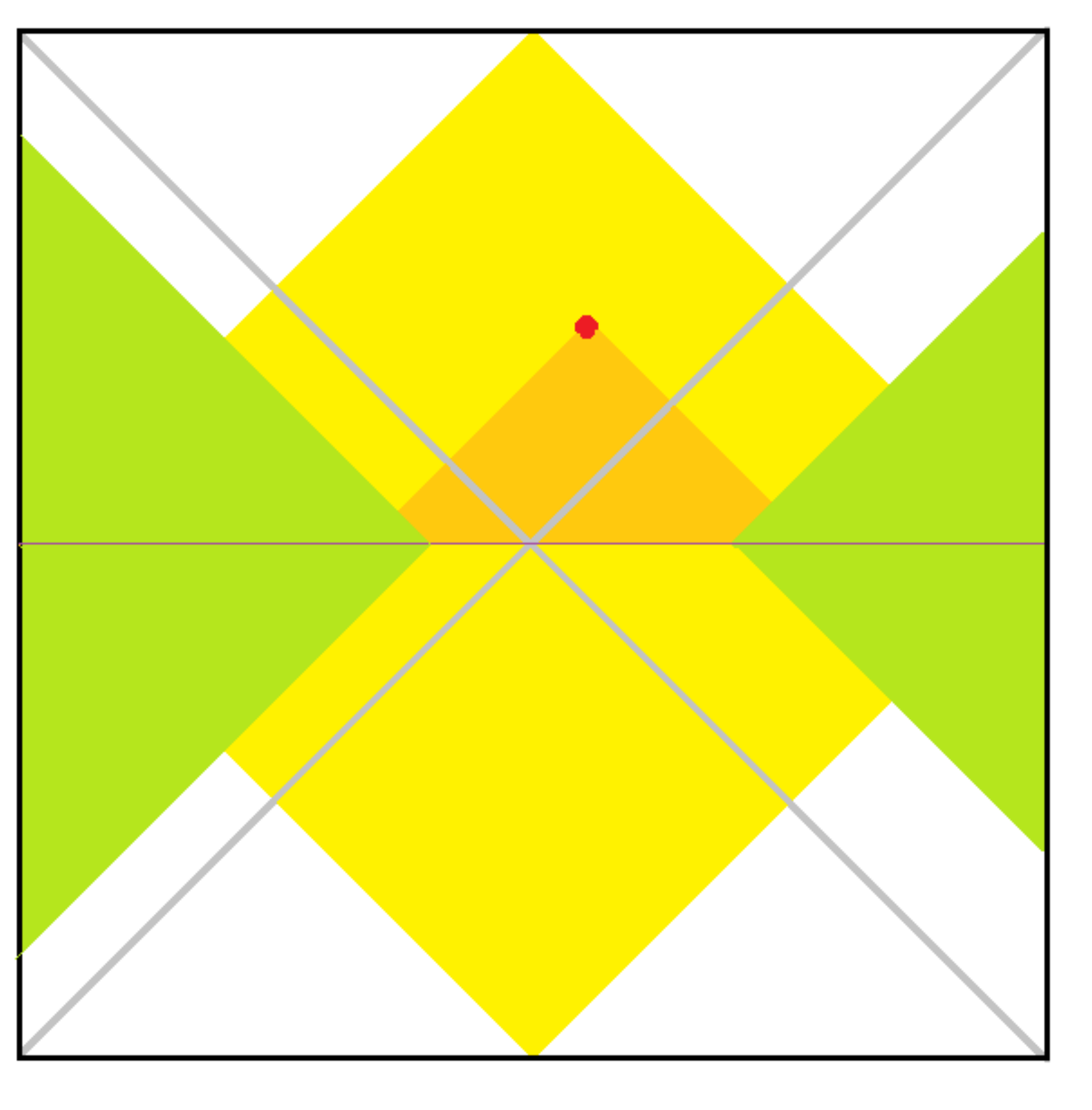}
\caption{The fields on the base $b$ may be translated to the boundaries by the HKLL method. }
\label{wdw3}
\end{center}
\end{figure}

The next step is to use the HKLL \cite{Hamilton:2005ju} construction to express the fields on the base of the triangle in terms of local gauge-invariant  CFT fields on the left and right boundaries. Figure \ref{wdw3} schematically shows the relevant regions for two specific points on the base $b$ of the orange triangle.

Two things are different for a field in the interior than for a field in either exterior region. The first is that both CFT's come into the construction. The other is that the bifurcate horizon is one of the points on  $b.$ According to our earlier arguments the field at that point has maximum complexity. Equivalently the entire range of boundary times comes into play when reconstructing $\phi(a).$  However, as long as ${\bf{a}}$ is not close to the bifurcate horizon, the contribution to the integral \ref{int-rep} from the immediate vicinity of the horizon will be negligible. Therefore in practice the relevant boundary times are a few ADS lengths. Nevertheless there is an approximation that would not be necessary for an exterior operator on either side.

\begin{figure}[h!]
\begin{center}
\includegraphics[scale=.3]{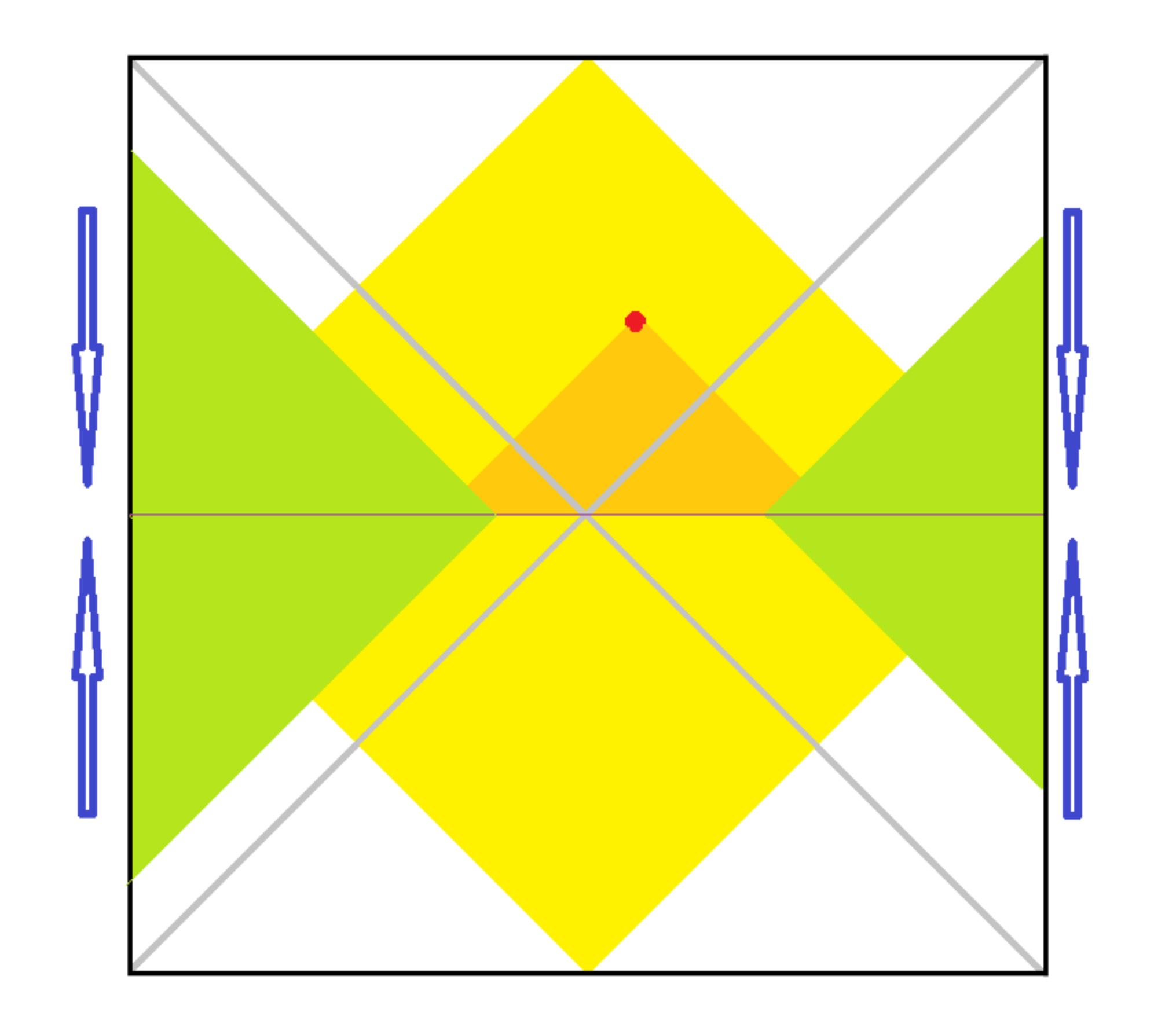}
\caption{ The final step in constructing $W_{LR}(a)$ is to run the boundary fields back to $t=0.$}
\label{wdw4}
\end{center}
\end{figure}

The final step is to use the gauge theory equations of motion to express the boundary fields in terms of gauge theory operators at $t=0.$ Schematically this is indicated in figure \ref{wdw4}.

The result of these operations is an expression for $\phi(a)$ as a non-local gauge theory operator involving both sides at $t=0.$  Let's call it $W_{LR}(a)$ where the notation implies that   $W_{LR}(a)$ is made of various kinds of  Wilson loops, and that $W_{LR}(a)$ is a function of both left and right CFT degrees of freedom. Under the stated restrictions the operator $W_{LR}(a)$ will be of relatively small complexity.

There are some obvious approximations in this procedure. All the non-linearities of the bulk equations of motion have been ignored. In particular this includes gravitational back reaction on the geometry. Perturbatively, such back reaction would be described by graviton loops, but under the low energy circumstances we have assumed, these corrections are negligible.

\begin{figure}[h!]
\begin{center}
\includegraphics[scale=.3]{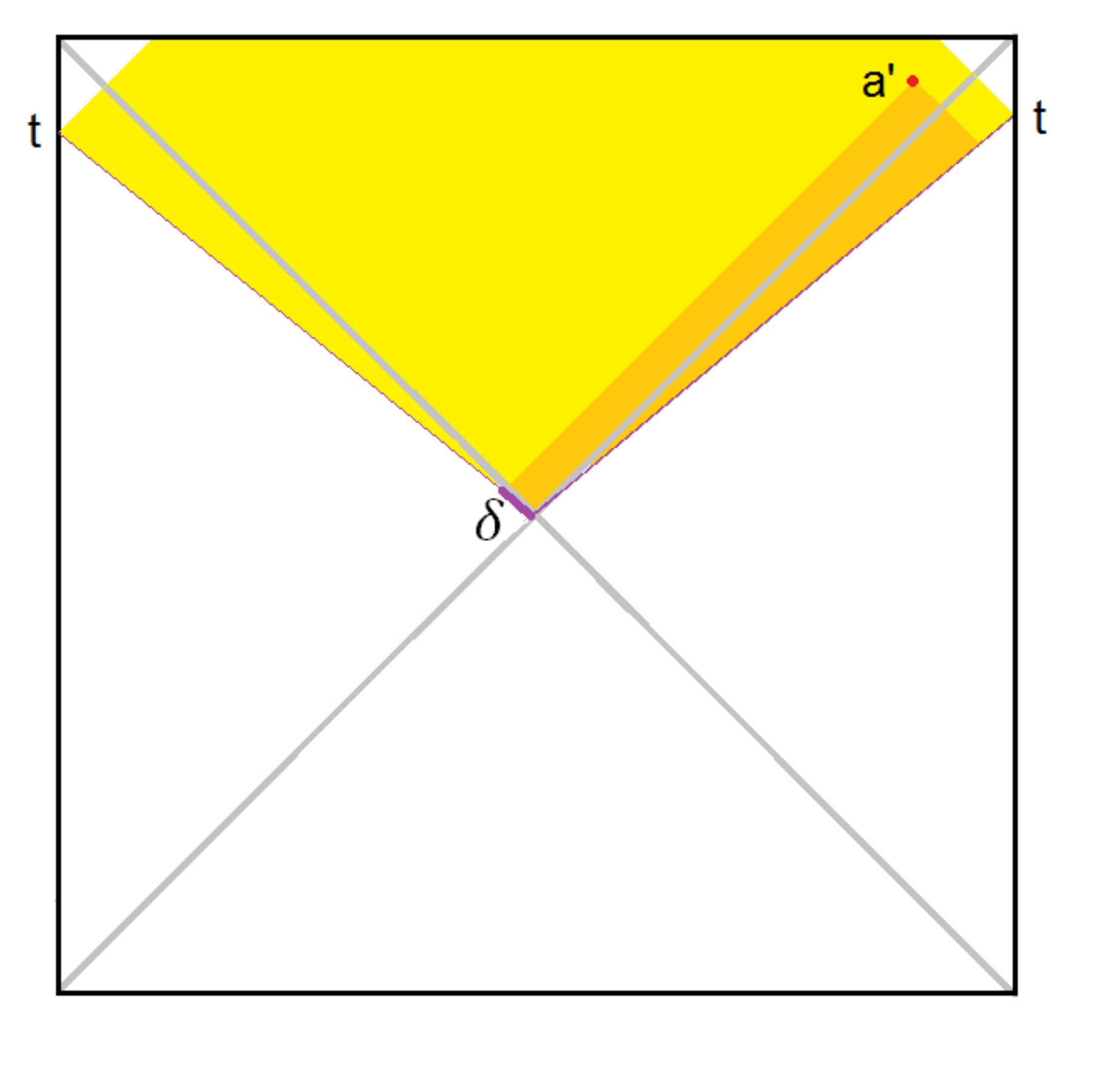}
\caption{The point  ${\bf{a}}$ may be pushed into the upper right corner of the Penrose diagram, where it would
be encountered by Bob if he falls in at a late time. The WDW patch is shown for $t_L=t_R.$ }
\label{wdw5}
\end{center}
\end{figure}

Now let us consider pushing the point ${\bf{a}}$ far into the future just behind Bob's horizon. The purpose is to predict Bob's experiences if he jumps in at a late time $t.$ The configuration is shown in the Penrose diagram as in figure \ref{wdw5}.
We can move   ${\bf{a}}$ to its new location  ${\bf{a}}'$  by boosting it by a large hyperbolic angle.

Our new goal is to represent $\phi(a')$ in the Schrodinger picture of the CFT's  on a boundary slice in which the time on both sides is Bob's jump-off time $t.$ In other words we would like to express $\phi(a')$ in the gauge theory representation of the Wheeler DeWitt patch in figure \ref{wdw5}. Both times, $t_L$ and $t_R$ need to be pushed ahead to $t.$

Let's shift the times in two steps. In the first step we shift $t_R$ ahead to $t$ and $t_L$ back to $-t.$ This is accomplished by acting with the boost operator
\be
U_-(t)= e^{-i (H_R -H_L) \ t}.
\label{boost }
\ee

\begin{figure}[h!]
\begin{center}
\includegraphics[scale=.3]{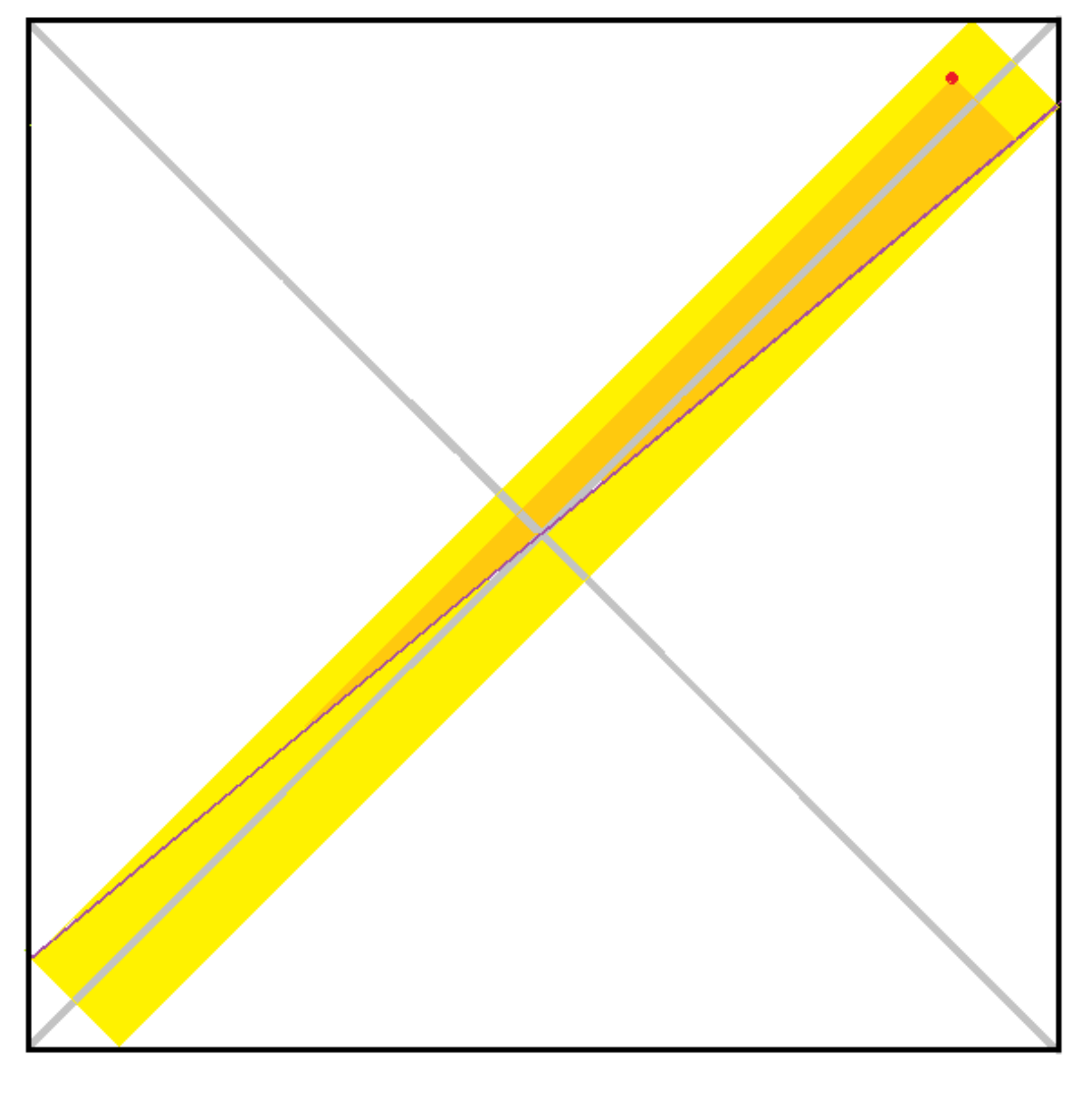}
\caption{The WDW patch for the case in which $t_L = -t_R.$    The diagram is related to figure \ref{wdw1} by a large boost.}
\label{wdw6}
\end{center}
\end{figure}

All this does is to boost figure \ref{wdw6} back to the original configuration of figure \ref{wdw1}. Thus after this first step
$\phi(a')$ has exactly the form $W_{LR}(a).$

In the second step we  push $t_L$ ahead by amount $2t$ until the times of both sides are equal to $t.$ The transformed description of $\phi(a')$ becomes,

\be
e^{-2it H_L }   W_{LR}(a)  e^{2it H_L }
\label{halfboost}
\ee

\bn
 At a formal level this gives an operator that acts on the two-sided wave function in the Schrodinger representation at time $t.$
The practical problem with this last procedure is that the operation in \ref{halfboost} introduces a great deal of complexity if $t$ is large. If $t>>t_{\ast} $ it pushes the complexity of the left dependence deep into the trans-scrambling region.

Being driven to trans-scrambling  complexity by \ref{halfboost}  must be reflected in sub-planckian distances somewhere in the geometry. Going back to
 figure \ref{wdw5}, we see a short line-segment labeled $\delta.$ As $t$ increases the proper length of this segment decreases. At $t=t_{\ast} $ we find $\delta$ has shrunken to the Planck length.

The bad news  of this section is that chaos and complexity make it impractical to compute the gauge theory description of even a simple low energy observable behind Bob's horizon at late time. The good news is that the difficulties are practical and do not present an in-principle obstacle to describing the interior.

\sc
\section{ Comments on Recurrences and Evaporation}

\subsection{Recurrences}
I've assumed that the classical geometric description the spacetime near Bob's end of the ERB is valid for arbitrarily late times. There are  reasons, described in Section 5.2 to think that the global geometry might break down by the classical recurrence time.
Certainly by the quantum recurrence time there must be a breakdown since the state of the two-sided system will return to something arbitrarily close to the TFD. That can't happens classically, for among other reasons, the stretching phenomenon discussed at the beginning of section 5 would not allow it.

One line of thought that I have not pursued  here suggests that if the TFD state is mildly perturbed, then Bob's horizon will remain smooth for a classical recurrence time, but then have a significant probability of fluctuating to a firewall\footnote{I thank Douglas Stanford for sharing his insights  on this topic.}. The argument is closely connected with both classical Poincare recurrences, and with the existence of a complexity maximum which is reached at the recurrence time.

 If such recurrent shockwaves do  occur they are manifestations of the dynamically closed nature of the ADS system. They are very reminiscent of Boltzmann fluctuations in  the causal patch of de Sitter space \cite{Dyson:2002pf}. In both cases, because the Hilbert space is finite all that can happen is that the system endlessly recycles, with statistical probability  governed by state-counting alone.
 In cosmology this leads to such undesirable and  counterintuitive behaviors as the dominance of Boltzmann Brains over normal brains, and the absence of an arrow of time. Firewalls may be similar phenomena.

 The solution to the cosmological problems is that cosmology does not take place in a finite Hilbert space. Because de Sitter space can decay to terminal ``hats" the system can leak out to an infinite Hilbert space. This may eliminate the  Boltzmann brain problem and provide an arrow of time \cite{Carroll:2004pn}\cite{Bousso:2011aa}\cite{Harlow:2011az}\cite{Susskind:2012pp}\cite{Guth}.

 ADS black holes are closed finite systems in which the Hawking radiation is bottled up. They also endlessly recycle and experience all  states with a statistical weight  independent of the starting point. Firewalls may occur just as Boltzmann Brains do in a closed cosmology. In this case the solution is to un-bottle the radiation. Evaporation causes the  system to leak out into an an infinite Hilbert space  in which recurrences don't happen.

Recurrences and their implications for the smoothness of horizons is poorly understood and will hopefully be the subject of future work.

\subsection{Evaporating Case  }

Computational complexity was first applied to black hole physics by Harlow and Hayden  \cite{Harlow:2013tf}. H-H were interested in proving that the basic AMPS experiment of distilling the bit $R_b$ from the Hawking radiation and then bringing it back into the black hole cannot be done. The issue of this paper is different; as in \cite{Maldacena:2013xja} we allow the possibility that the experiment can be done, and that it does indeed create a disturbance behind the horizon; the mechanism being transmission through the ERB. But that  only implies a disturbance \it  if the experiment were done \rm.

The question is whether sending a firewall through the ERB is so easy that it is likely to happen by generic interaction of the radiation with the environment. This is very unlikely; the obstacle to sending a message through the ERB is much larger in the evaporating case than in the eternal ADS case, where it is already big. The argument is based on the technical  results of Harlow and Hayden.

In applying the ideas of computational complexity to the difficulty of sending messages, there is an important difference between the two-sided ADS case and the evaporating black hole. Let us consider a schematic simplification of the two cases. The simplified version of the ADS  replaces the initial TFD state by  a product of $K$ bell pairs, each pair being shared between Alice and Bob.

\be
|TFD\ra \to \left\{ \ |00\ra + |11\ra \ \right \}^{\otimes K}
\ee

The time evolution by a time $t$ is given by applying a left and right evolution operator.

\be
|\psi(t)\ra = U_L(t) U_R(t) \left\{ \ |00\ra + |11\ra \ \right \}^{\otimes K}
\ee

Now consider a left precursor

\be
W_p = U_L(t) W U_L(t)^{\dag}
\label{conj}
\ee

The complexity of $W_p$ is of the same order as the complexity of $U_L$ which is of order $t.$ In general we are interested in values of $t$ that are polynomial in the entropy of the black hole.  In practical terms that is a lot of complexity, but compared to the maximum value it is  small.
In coming to this conclusion, namely that the complexity of $W_p$ is of order $t$ it is essential that the two subsystems do not interact  and transfer information during the evolution.

Now consider the evaporating case after the Page time. The model that has been widely used is to divide the system into $B$ black hole qubits, and $R$ radiation qubits. However, unlike the ADS case the two subsystems do not evolve separately. There is continuous transfer of degrees of freedom from the black hole to the radiation.

A simple  model is to take $R+B$ qubits in a fully scrambled state. The scrambled state can be obtained by applying a unitary $U_{R+B}$ in the $R+B$ system. Once scrambled the two subsystems  can be separated into radiation and the remaining black hole.

Since the system has evolved for a polynomial amount of time the scrambling unitary operator  will be of polynomial complexity. One can undo it in polynomial time and find a precursor for any initial degree of freedom. However, since the two systems, $R$ and $B,$ have interacted, this operation must involve both the radiation and the black hole degrees of freedom. In the ADS case constructing a precursor only involved Alice's degrees of freedom (the left-side CFT).

Nevertheless, even if Alice only has control  over the  radiation, she can still isolate a degree of freedom  $R_b$ that is entangled with any given qubit in the $B$ subsystem.  If one considers the code subspace--- a subspace of the $R$ Hilbert space--- which is entangled with $B,$ there is a unitary operator $V$ which transforms the entangled state to a product of Bell pairs. $V$ is an operator made only of the radiation degrees of freedom.

By using $V$ as a replacement for $U_L$ in \ref{conj} Alice can construct the precursor, but the operator $V$ is far more complex than either $U_L$ or $U_{R+B}.$ Both of these operators were ordinary evolution operators for polynomial time. $V$ is not an evolution operator, and
the result of Harlow and Hayden shows that $V$ is maximally complex with a complexity of order $e^R.$

An intuitive explanation for the large complexity of $V$ might go like this: It would be relatively easy for Alice if she had access to both $R$ and $B$ cubits; then she could use $U_{R+B}$ in \ref{conj}. But she is handicapped by not being allowed to employ the qubits in $B.$ At first sight her task might seem totally impossible. But because the space of states is finite, it is guaranteed that the complexity of $V $ is bounded from above  by $e^R.$ Harlow and Hayden show that it is not likely to be smaller than $e^R.$

The  result implies that it takes an exponential time to distill $R_b,$ by which time the black hole has long evaporated. This may be so, but even if Alice could find some way of miraculously speeding up her quantum computer, she still has to do an extraordinarily  complex operation to send a signal through the ERB. It's not likely that such operations occur by chance interaction with the environment.

 \sc
 \section{Conclusion}

Firewalls, like reversals of the second law,  can occur in one of  two ways; either by the action of an extremely complex operation needed to  locally reverse the arrow of time, or by
 bottling up the system for a classical recurrence time. Temporary firewalls can also form if black holes are formed over long time periods. The argument in Section 7 suggests that such firewalls last for a time no longer than the formation time.

 In the two-sided ADS case the  complexity required to send a firewall through an ERB is large, but assuming Alice acts after a polynomial time, the complexity will only be polynomial. In practice that is a large complexity, not likely to be encountered naturally, but still, it is very small compared to the maximum.

 On the other hand if the radiation is not bottled, that is, if it is free-streaming into empty space, the situation is much less favorable for firewall production. The Harlow-Hayden calculation shows that the complexity of the operator $V$ that Alice must act with, is maximal; i.e., exponential in the entropy. Moreover, if the radiation is un-bottled
 recurrences do not occur.
It is ironic that evaporation may be the ingredient which  eliminates the possibility of firewalls.

We also touched on another issue in Section 5; namely the properties of a typical black hole. The concept of typical depends on the ensemble whose members we choose to include. Haar-averaging over all states in a given energy range defines a possible ensemble but it is inappropriate for modeling real black holes formed by collapse. For those we should restrict to the states  satisfying the following criteria:

\bn
1) The initial horizon should be smooth and the state relatively simple.

\bn
2) The interior should be stretching and not contracting.

\bn
3) The complexity of the state should be far below the maximum value. This last condition, which applies to the bottled up case, would follow if we require the time to be much less than the classical recurrence time.

\bigskip

It is true that these restrictions eliminate almost all states because almost all states have close to maximal complexity \cite{Knill}.
But these conditions seem plausible for representing the conditions under which real black holes are created.
With this ensemble  the evidence seems strong that  firewalls do not typically exist, unless Alice does an exceedingly complex operation; one which is very unlikely to occur by natural processes.

The connection between complexity and the geometry of the near-horizon region may be of wider interest than defusing the firewall argument.  It may illuminate problems such as the meaning of sub-planckian distances. In addition there appear to be interesting connections between complexity and classical gravitational phenomena. One example is the gravitational force in the near-horizon region. A second is the relation between the growth of complexity of a quantum state and the stretching of the interior of a black hole.
It is  possible that these connections  may shed light on problems of complexity by mapping them to gravity problems.

\appendix

\section{The Meaning of Small Changes}

I referred a number of times to small changes in the state of a system.

In quantum mechanics a state is a vector that evolves by linear unitary mapping. Two states which start close in Hilbert space stay close as a consequence of the time independence of the inner product. Therefore one does not judge the chaotic nature of a system by whether quantum states exponentially diverge: they don't.

By a small change I mean one which takes a state to an orthogonal state in which very few degrees of freedom have been changed.

In a quantum circuit one might act on the state with a unitary operator close to the identity. This will generally carry the quantum state to a quantum-mechanically nearby   state, and the evolution of the circuit will preserve the quantum-nearness no matter how long it evolves.

By contrast if we take $W$ to be a traceless Pauli operator acting on a single qubit, then for the overwhelming majority of states it leads to an orthogonal state but one in which only a single degree of freedom has been affected. This is what I mean by a small change.

\section*{Acknowledgements}

I am grateful to Patrick Hayden, Stephen Shenker, and Douglas Stanford for numerous discussions.

Support for this research came through NSF grant Phy-1316699 and the Stanford Institute for Theoretical Physics.

\end{document}